\documentclass[aps,superscriptaddress,10pt,nofootinbib,notitlepage,twocolumn]{revtex4-1}

\usepackage[normalem]{ulem}

\usepackage[latin1]{inputenc}
\usepackage{graphicx}
\usepackage{float}
\usepackage{latexsym}
\usepackage{graphicx}
\usepackage{amssymb}
\usepackage{amsmath}
\usepackage{amsfonts}
\usepackage[colorlinks=true,citecolor=blue,hyperfootnotes=false]{hyperref}
\usepackage[dvipsnames]{xcolor}

\usepackage{cool}
\usepackage{dcolumn}
\usepackage{textcomp}
\usepackage{xfrac}
\usepackage{slashed}
\usepackage{multirow}

\def\be{\begin{equation}}
\def\ee{\end{equation}}
\def\figs/B{B}
\def\bea{\begin{eqnarray}}
\def\eea{\end{eqnarray}}
\def\bg{\begin{eqnarray}}
\def\nd{\end{eqnarray}}

\def\cos{{\rm cos}}

\def\log{{\rm log}}

\def\be{\begin{equation}}
\def\ee{\end{equation}}

\begin{document}

\title{Nonminimal Couplings and the Forgotten Field of Axion Inflation }

\author{Evan McDonough}

\affiliation{Center for Theoretical Physics, Laboratory for Nuclear Science, and Department of Physics,
Massachusetts Institute of Technology, Cambridge, MA 02139, USA}%

\affiliation{Kavli Institute for Cosmological Physics and Enrico Fermi Institute, The University of Chicago, Chicago, IL 60637, USA}

\affiliation{Department of Physics, University of Winnipeg, Winnipeg MB, R3B 2E9, Canada}

\author{Alan H.~Guth}
\affiliation{Center for Theoretical Physics, Laboratory for Nuclear Science, and Department of Physics,
Massachusetts Institute of Technology, Cambridge, MA 02139, USA}

\author{David I.~Kaiser}
\affiliation{Center for Theoretical Physics, Laboratory for Nuclear Science, and Department of Physics,
Massachusetts Institute of Technology, Cambridge, MA 02139, USA}

\begin{abstract}
We study the multifield dynamics of axion models nonminimally coupled to gravity.  As usual, we consider a canonical $U(1)$ symmetry-breaking model in which the axion is the phase of a complex scalar field. If the complex scalar field has a nonminimal coupling to gravity, then the (oft-forgotten) radial component can drive a phase of inflation prior to an inflationary phase driven by the axion field. In this setup, the mass of the axion field is dependent on the radial field because of the nonminimal coupling, and the axion remains extremely light during the phase of radial inflation. As the radial field approaches the minimum of its potential, there is a transition to a second phase of inflation during which the axion field contributes a substantial fraction of the system's total energy density, yielding a phase of ``natural inflation.'' This system exhibits ultra-light isocurvature perturbations, which are converted to adiabatic perturbations by a turning field-space trajectory. For models wherein the CMB pivot scale exited the horizon well before the turn, this acts to suppress the tensor-to-scalar ratio $r$, without generating CMB non-Gaussianity or observable isocurvature perturbations. Finally, we note that for certain parameters the interaction strength between axion and gauge fields can be suppressed during the first phase of inflation relative to its value during the second phase by several orders of magnitude. This decouples the constraints on the inflationary production of gauge fields (e.g., from primordial black holes) from the constraints on their production during (p)reheating.
\end{abstract}

\date{June 6, 2024}

\maketitle

\section{Introduction}
\label{sec:intro}

The axion field was proposed as a solution to the strong CP problem of the Standard Model of particle physics \cite{Peccei:1977hh,Wilczek:1977pj,Weinberg:1977ma}. More recently, axions rose to prominence in early universe cosmology both as a candidate for the inflaton field \cite{Freese:1990rb,McAllister:2008hb,Silverstein:2008sg} and as a candidate for cold dark matter \cite{Preskill:1982cy,Abbott:1982af,Dine:1982ah}, bolstered by interesting phenomenology of the characteristic axion coupling to gauge fields (e.g., Ref.~\cite{Barnaby:2011qe}). These developments coincided with the discovery of flux compactifications of string theory, which generically include hundreds of axion fields \cite{Svrcek:2006yi,Arvanitaki:2009fg,Cicoli:2012sz}. Multi-axion (`axiverse') models can also arise directly from field theory \cite{Maleknejad:2022gyf,Alexander:2023wgk,Alexander:2024nvi}.

As the Goldstone boson of a spontaneously broken (approximate) global $U(1)$ symmetry, the axion is naturally described as the phase of a complex scalar field. Inherent in this construction is a second scalar field, the radial component, which plays the role of the order-parameter of the symmetry breaking. Thus, axion models, both as they were initially conceived \cite{Peccei:1977hh,Wilczek:1977pj,Weinberg:1977ma}, and in their modern incarnations, are inherently multifield theories. As such, the dynamics of such models during the early universe should be studied using techniques developed in recent years for the analysis of multifield inflation, with a focus on characteristically multifield phenomena such as isocurvature perturbations, primordial non-Gaussianity, and their compatibility with recent observations. (For reviews, see Refs.~\cite{Wands:2007bd,Chen:2010xka,Gong:2016qmq}.)

Beginning nearly a decade prior to the proposal of the axion, physicists began to clarify that self-interacting scalar fields in curved spacetime will generically develop nonminimal couplings to gravity \cite{Chernikov:1968zm,Callan:1970ze,Bunch:1980br,Bunch:1980bs,Birrell:1982ix,Odintsov:1990mt,Buchbinder:1992rb,Parker:2009uva,Markkanen:2013nwa}. Even if the dimensionless nonminimal coupling constants $\xi$ vanish at tree-level in a given model, they will be generated by loop corrections, and hence they are required for self-consistent renormalization of scalar-field models in curved spacetime. From the perspective of effective field theory (setting renormalization aside), in $(3 + 1)$ spacetime dimensions such couplings take the form of operators in the action with mass dimension 4, and hence they should be included unless forbidden by the specific symmetries of a given theory.

In this work we study multifield inflation in the simplest axion model, including the nonminimal coupling of the radial field to gravity. The potential for the axion field, when expressed in the Einstein frame, is exponentially sensitive to the radial field. This leads to dynamics akin to hybrid inflation \cite{Linde:1991km,Linde:1993cn,Copeland:1994vg}, with a phase of inflation driven by the radial field followed by a second phase of inflation driven predominantly by the axion or by an admixture of the two fields, depending on the parameters on the model. This bears a strong resemblance to dynamics recently observed in supergravity \cite{Linde:2018hmx,Aldabergenov:2020bpt} and supergravity-inspired \cite{Christodoulidis:2018qdw} multifield models. It is particularly striking that the behavior of this model emerges from such simple ingredients.

Turning to cosmological perturbations, we find a phenomenon recently observed in multifield models with curved field-space manifolds \cite{GrootNibbelink:2000vx,GrootNibbelink:2001qt,Seery:2005gb,Langlois:2008mn,Peterson:2010np,Gong:2011uw,Kaiser:2012ak,Renaux-Petel:2015mga,Gong:2016qmq,Christodoulidis:2018qdw,Christodoulidis:2019mkj,Christodoulidis:2019jsx,Aldabergenov:2020bpt,Fumagalli:2020adf,Grocholski:2019mot,Garcia-Saenz:2018ifx,Garcia-Saenz:2018vqf,Fumagalli:2019noh,Garcia-Saenz:2019njm,Pinol:2018euk,Pinol:2020cdp,Fumagalli:2019ohr,Ashoorioon:2019kcy}, namely, ultra-light isocurvature perturbations \cite{Palma:2020ejf}. In the model developed here, the isocurvature perturbations are ultralight during the first inflation phase, and develop a nearly scale-invariant spectrum of perturbations on large scales. They become super-heavy during the second inflation phase, and subsequently rapidly decay. At the interface of these two regimes, there can be a rapid conversion of isocurvature into curvature perturbations, leading to an overall enhancement of the amplitude of the curvature perturbation power spectrum. In the language of multifield inflation, this system exhibits a turning trajectory with a fast turn. 

This has important implications for next-generation cosmic microwave background (CMB) experiments, in particular regarding the observation or non-observation of primordial gravitational waves, since the tensor-to-scalar ratio $r$ can be suppressed in this scenario due to the relative enhancement of the scalar spectrum. We find that the suppression of $r$ can become non-negligible even for modest values of the nonminimal coupling, $\xi \gtrsim {\cal O} (0.2)$. This builds on Refs.~\cite{Achucarro:2015caa,Alam:2024krt,Montefalcone:2022jfw,Salvio:2021lka,Salvio:2023cry} that have shown that Natural Inflation can be compatible with Planck data once multifield dynamics are included\footnote{Natural Inflation can also be compatible with Planck data in the context of modified gravity Modified\cite{Racioppi:2024zva,Salvio:2019wcp,Salvio:2022mld} or alternatives to $\Lambda$CDM model\cite{RoyChoudhury:2022rva,Bostan:2023ped}.}.

Finally, this scenario has implications for the phenomenology of axions. Considering the axions' canonical interaction with gauge fields, we find the interaction strength during the radial phase can be suppressed relative to the interaction strength during the second inflation phase. This decouples the constraints on inflationary production of gauge fields (e.g., from primordial black holes) from the constraints on production at (p)reheating.

The remainder of this paper is organized as follows. In Sec.~\ref{sec:nonminimal} we introduce the model and identify parameter ranges of interest. We analyze the inflationary background dynamics in Sec.~\ref{sec:inflation}, and study the evolution of perturbations in Sec.~\ref{sec:perts}. We compute the associated observables and constraints in Sec.~\ref{sec:observables}, and in Sec.~\ref{sec:gaugefields} we perform a brief analysis of the interaction with gauge fields. We close in Sec.~\ref{sec:discussion} with a discussion of directions for future work.

\section{Nonminimal Couplings and Axions}
\label{sec:nonminimal}

The starting point for this work is the operational definition of the axion field: the pseudo-Nambu-Goldstone boson of a spontaneously broken global $U$(1) symmetry, which gains a mass through an explicit symmetry breaking by nonperturbative effects. (See, e.g., Refs.~\cite{Peccei:1977hh,Wilczek:1977pj,Weinberg:1977ma} for the original works, and Refs.~\cite{Svrcek:2006yi,Arvanitaki:2009fg,Cicoli:2012sz} for ubiquitous realizations in string theory.) This theory is naturally described in terms of a complex scalar field $\Phi$, with the axion $\vartheta$ defined as the phase of $\Phi$: $\Phi=\varphi e^{i\vartheta}$, where $\varphi$ and $\vartheta$ are real-valued scalar fields.

Self-interacting scalar fields in curved spacetime will generically develop a nonminimal coupling to gravity, acquired through loop effects \cite{Chernikov:1968zm,Callan:1970ze,Bunch:1980br,Bunch:1980bs,Birrell:1982ix,Odintsov:1990mt,Buchbinder:1992rb,Parker:2009uva,Markkanen:2013nwa}. More generally, the interaction $|\Phi|^2 R$ is a dimension-4 operator allowed by the symmetries of the problem, and thus, from the perspective of effective field theory, must be included. With this in mind, in this work we generalize the canonical axion model to incorporate a nonminimal coupling of the complex scalar field $\Phi$ to gravity. 

We work in $(3 + 1)$ spacetime dimensions and consider an action of the form
\begin{equation}
    S_J = \int d^4 x \sqrt{- \tilde{g}} \bigg[  f (\Phi) \tilde{R} - \frac{1}{2} \vert \partial \Phi \vert^2 - V (\Phi) \bigg] \, .
\label{eq:axionNonMinimal}
\end{equation}
The subscript $J$ denotes that the action of Eq.~(\ref{eq:axionNonMinimal}) is written in the Jordan frame, in which the nonminimal coupling $f (\Phi) \tilde{R}$ remains explicit. Given that the axion, by construction, is not subject to perturbative breaking of the $U(1)$ symmetry, the nonminimal coupling is constrained to depend only on $|\Phi| \equiv \varphi$.

The potential for $\Phi$ consists of two contributions. The spontaneous symmetry breaking (``Higgs'') potential, $V_{\rm SSB} = \lambda \left( |\Phi|^2 - v^2 \right)^2$/4, ensures that the ground state of the theory is the $U$(1) symmetry-breaking state $\langle |\Phi| \rangle =v$. In addition, nonperturbative effects generate a potential for the axion field, $V_{a} = \Lambda^4 (1-\cos\, \vartheta)$, as in ``natural inflation" \cite{Adams:1992bn}. The value of $\Lambda$ is determined by the microphysics of the theory. In QCD constructions, e.g., the Peccei-Quinn axion \cite{Peccei:1977hh}, $\Lambda^4 = \chi$ is the topological susceptibility of the QCD vacuum (for a review, see, e.g., Ref.~\cite{PhysRevD.98.030001}). More generally, one might expect $\Lambda$ to be a function of the radial field $\varphi$; indeed, in string theory constructions this is generally the case (see, e.g., Ref.~\cite{Cicoli:2012sz}). For simplicity, we will take $\Lambda$ to be a constant and a free parameter, constrained only by the requirement that the energy scale of the explicit $U(1)$ symmetry breaking ($\Lambda$) be lower than the energy scale of spontaneous $U(1)$ symmetry breaking ($v$).

To simplify the analysis, we decompose $\Phi$ into real fields as $\Phi=\varphi e^{i\vartheta}$. The action then reads
\begin{equation}
    \begin{split}
        S_J = \int d^4 x \sqrt{ - \tilde{g}} &\bigg[ f (\varphi) \tilde{R} - \frac{1}{2} \left( \partial \varphi \right)^2 - \frac{1}{2} \varphi^2 \left( \partial \vartheta \right)^2 \\
        &\> - \frac{ \lambda}{4} \left( \varphi^2 - v^2 \right)^2 - \Lambda^4 \left( 1- \cos \vartheta \right) \bigg] \, .
    \end{split}
    \label{SJ2}
\end{equation}
The kinetic terms are those of conventional polar coordinates on a flat field space. This defines a metric on the field-space manifold in the Jordan frame, ${\cal G}_{IJ} ^{(J)}$, with components,
\begin{equation}
    {\cal G}_{\varphi \varphi} ^{(J)} = 1 \;\; , \;\;  {\cal G}_{\vartheta \vartheta} ^{(J)} = \varphi^2 ,
\end{equation}
where the superscript $J$ refers to Jordan frame.

We next perform a conformal transformation of the spacetime metric, to work in the Einstein frame \cite{Kaiser:2010ps,Abedi:2014mka}
\begin{equation}
    g_{\mu \nu} = \frac{ 2 f (\varphi) }{M_{\rm pl}^2} \tilde{g}_{\mu \nu}
    ,
\end{equation}
where $M_{\rm pl}$ is the reduced Planck mass, $1/\sqrt{8 \pi G}$. This serves to make the coupling to gravity canonical; in doing so it modifies both the kinetic and potential terms in the action. The action in the Einstein frame takes the form
\begin{equation}
    S_E = \int d^4 x \sqrt{-g} \bigg[ \frac{ M_{\rm pl}^2}{2} R - \frac{1}{2} g^{\mu\nu} {\cal G}_{IJ}^{(E)} \partial_\mu \phi^I \partial_\nu \phi^J - V_E (\phi^I ) \bigg] \, ,
\label{SE}
\end{equation}
for $\phi^I = (\varphi, \vartheta)$. The field-space metric in the Einstein frame becomes
\begin{equation}
    {\cal G}_{IJ} ^{(E)} = \frac{M_{\rm pl}^2 }{ 2 f} {\cal G}_{IJ} ^{(J)} + \frac{3 M_{\rm pl}^2}{2}\frac{f_{,I} f_{,J}}{f^2} \, ,
\end{equation}
where $f_{,I} \equiv \partial f / \partial \phi^I$ denotes a derivative with respect to field $\phi^I$. In our simple case, the field-space metric remains diagonal, with the nonvanishing components
\begin{equation}
    {\cal G}_{\varphi \varphi}^{(E)} = \frac{ M_{\rm pl}^2}{2f} \left( 1 +  \frac{ 3 f_{, \varphi}^2}{f} \right) \>\> , \>\> {\cal G}_{\vartheta\vartheta}^{(E)} = \frac{ M_{\rm pl}^2}{2f} \varphi^2 \, .
    \label{GEIJ}
\end{equation}
The kinetic term of the axion defines the axion decay constant in the Einstein frame,
\begin{equation}
\label{eq:fa}
    f_a = \frac{\langle \varphi \rangle}{\sqrt{2 f}} M_{\rm pl} \, .
\end{equation}
Meanwhile the potential is rescaled, and is given by
\begin{equation}
\label{eq:VE1}
    V_E (\varphi, \vartheta) = M_{\rm pl}^4 \frac{V (\varphi, \vartheta) }{4 f^2 (\varphi)}.
\end{equation}
We consider the usual form of the nonminimal coupling \cite{Chernikov:1968zm,Callan:1970ze,Bunch:1980br,Bunch:1980bs,Birrell:1982ix,Odintsov:1990mt,Buchbinder:1992rb,Parker:2009uva,Markkanen:2013nwa},
\be
f(\varphi) = \frac{1}{2}\left( M^2 + \xi \varphi^2 \right) \, .
\label{fdef}
\ee
To remain consistent with late-time observations, we require $2f (v) = M_{\rm pl}^2$ when $\varphi$ reaches the minimum of its potential, $\langle \varphi \rangle = v$. This in turn requires $M^2 = M_{\rm pl}^2 - \xi v^2$. We note that in the ground state, $\langle \varphi \rangle = v$, the axion decay constant $f_a$ of Eq.~(\ref{eq:fa}) obeys $f_a = v$, independent of $\xi$ and $M$.

The conformal transformation to the Einstein frame transforms the flat field space of the Jordan frame into a {\it curved} field space \cite{Kaiser:2010ps}. The Ricci scalar of the field-space manifold in the Einstein frame is a function of the radial field $\varphi$, as shown in Fig.~\ref{fig:Ricci}. The full form is given in Appendix \ref{app:fieldspace}. The curvature is peaked at $\varphi=0$, where it takes a value set by $\xi$: ${\cal R}_E= 4\xi(1+3\xi)/M_{\rm pl}^2$. More generally, one can identify two regimes of interest. To do so, we define the the ratio,
\begin{equation}
\label{eq:rvarphi}
    r_{\varphi} \equiv \frac{\sqrt{\xi} \varphi}{M}.
\end{equation}
For $r_{\varphi}\ll 1$, the curvature can be expanded as
\be
\label{eq:REsmallphi}
\mathcal{R}_{E}|_{r_{\varphi}\ll 1}=\frac{4 \xi (1+3\xi)}{M_{\rm pl}^2}\left(1 + \mathcal{O}(r_{\varphi}^2) \right) ,
\ee
whereas in the opposite regime, $r_{\varphi} \gg 1$, it takes the form
\be
\label{eq:RElargephi}
\mathcal{R}_{E}|_{r_{\varphi} \gg 1} = \frac{4 \xi}{M_{\rm pl}^2 (1+ 6\xi) 
r_{\varphi}^2}\left(1 +
\mathcal{O}\left(\frac{1}{r_{\varphi}^2} \right) \right).
\ee
In the limit $\sqrt{ \xi} \, v\gg M$, $\xi$ held fixed, the vacuum is well approximated by a flat field space ($\mathcal{R}_{E}\ll 1/M_{\rm pl}^2$), whereas for $\sqrt{\xi} \, v \ll M$, the field space can be strongly curved.

\begin{figure}[h!]
\includegraphics[width=0.49\textwidth]{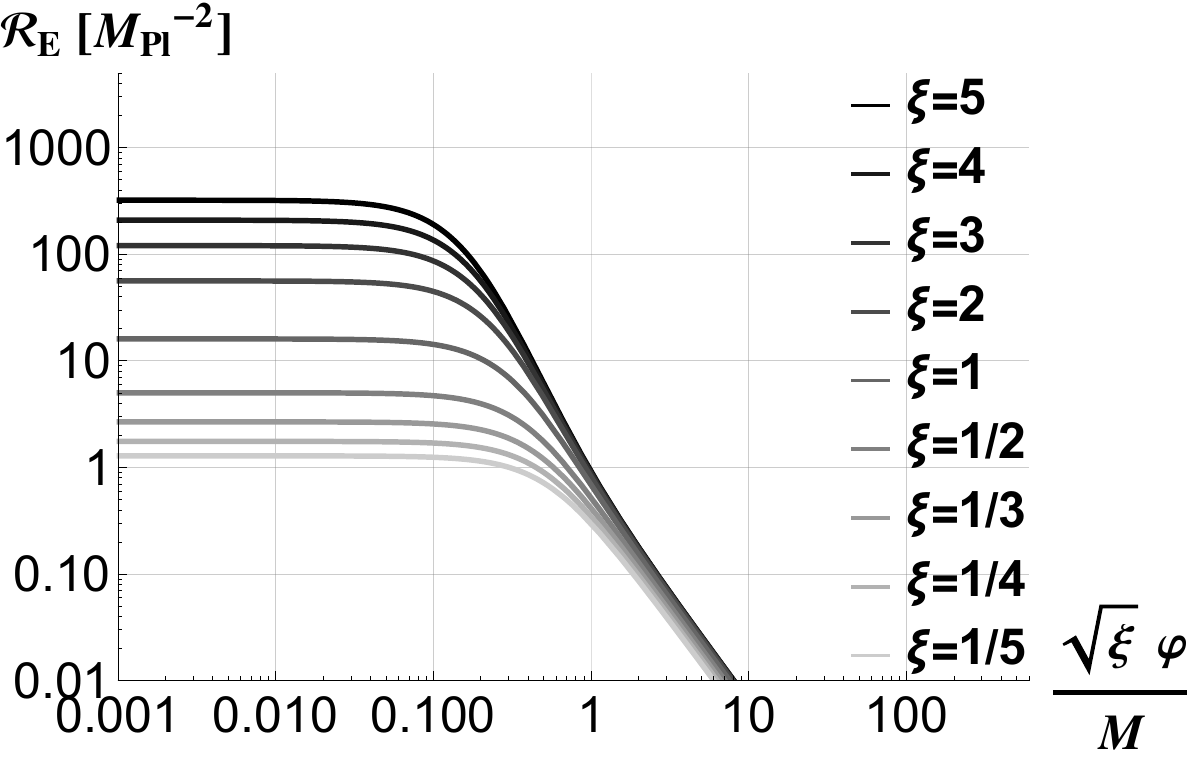}
\caption{The Ricci scalar of the field-space manifold in the Einstein frame, ${\cal R}_E$. This quantity depends only on the radial field $\varphi$, and takes the form ${\cal R}_E =F(\xi, r_{\varphi})/M_{\rm pl}^2 $ with $r_{\varphi}=\sqrt{\xi}\varphi/M$.
The curvature at $\sqrt{\xi}\varphi/M \ll 1$ is set by $\xi$; in the figure we show ${\cal R}_E$ for $\xi=1/5$, $1/4$, $1/3$, $1/2$, $1$, $2$, ..., $5$ (bottom to top). }
\label{fig:Ricci}
\end{figure}

The background dynamics follow from the equations of motion for the fields and the Friedmann equation. These can be written,
\bea
&& \ddot \varphi + 3 H \dot \varphi + \Gamma^{\varphi} _{\varphi \varphi} \dot \varphi^2 + \Gamma^{\varphi} _{\vartheta\vartheta} \dot \vartheta^2 + {\cal G}^{\varphi \varphi} V_{E,\varphi}=0\\ && \ddot \vartheta + 3 H \dot \vartheta + 2\,\Gamma^{\vartheta} _{\vartheta \varphi} \dot \varphi \dot{\vartheta} + {\cal G}^{\vartheta \vartheta}V_{E,\vartheta}=0, \label{eomaxion}
\eea
where $\Gamma^{I} _{JK}$ denote the Christoffel symbols associated with the field-space metric ${\cal G}_{IJ}^{(E)}$, given in Appendix \ref{app:fieldspace}. The Friedmann equation takes the form
\bea
3 M_{\rm pl}^2 H^2 = \frac{1}{2} && \left(\frac{M_{\rm pl}^2}{2f} \right) \left[1 + \frac{6 \xi^2 \varphi^2}{2f} \right] \dot{\varphi}^2 \\ &&+
\frac{1}{2} \left(\frac{M_{\rm pl}^2}{2f} \right) \varphi^2  \dot{\vartheta}^2 + V_{E} , \nonumber
\eea
where the Einstein-frame potential is given by
\be
\hspace{-0.2cm}V_E =\frac{\lambda M_{\rm pl}^4}{4}  \frac{ (\varphi^2 - v^2)^2}{( M^2 + \xi \varphi^2)^2}  +  \frac{ M_{\rm pl}^4\Lambda^4 }{( M^2 + \xi \varphi^2)^2}\left( 1 - \cos \vartheta\right),
\label{eq:VE}
\ee
as per Eq.~\eqref{eq:VE1}. One can appreciate from Eq.~\eqref{eq:VE} that the change to the Einstein frame, at large $\varphi$, flattens the symmetry-breaking potential and suppresses the magnitude of the axion potential.

 \begin{figure*}[t]
\centering
\includegraphics[width=0.49\textwidth]{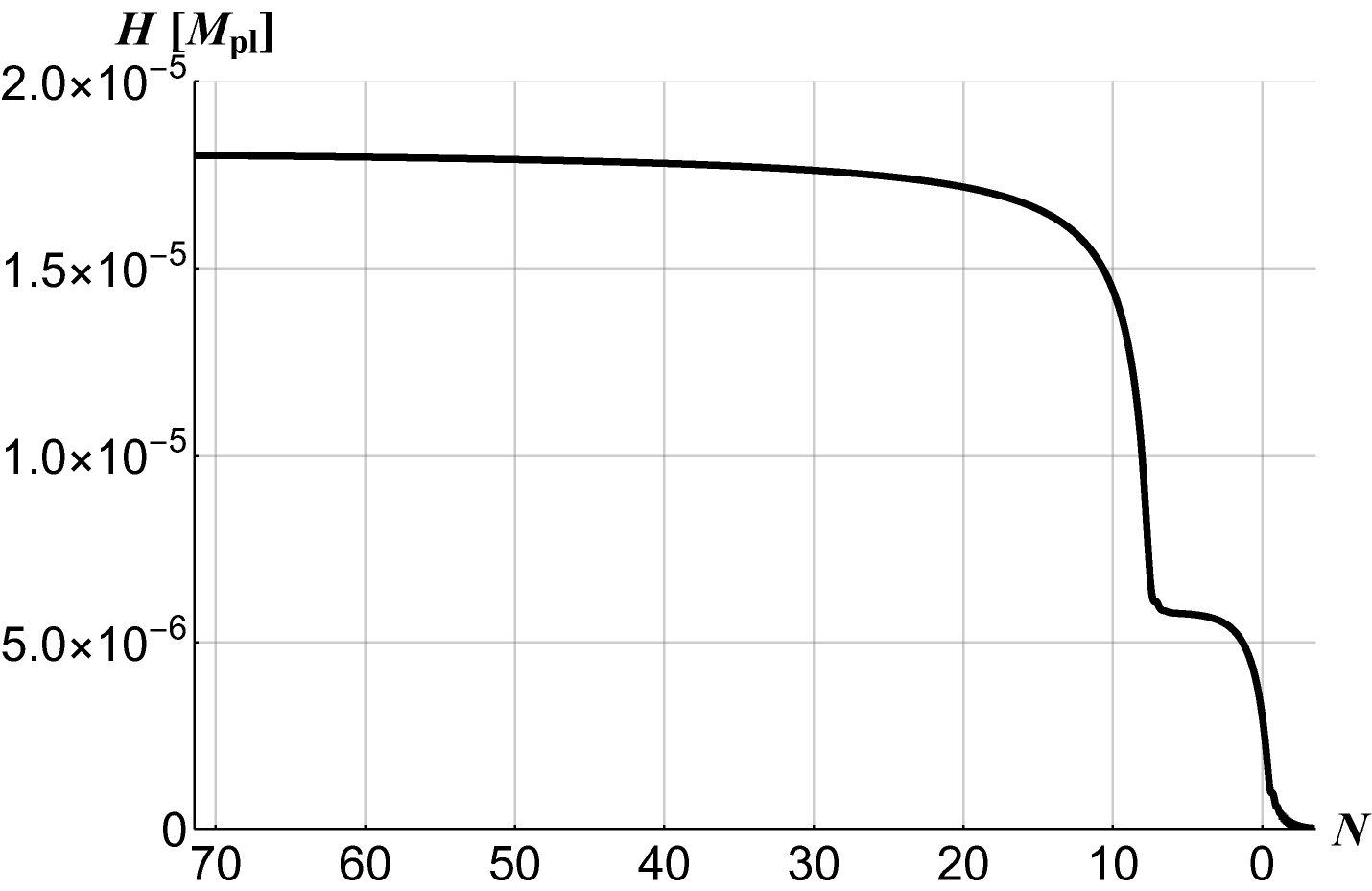}
\includegraphics[width=0.49\textwidth]{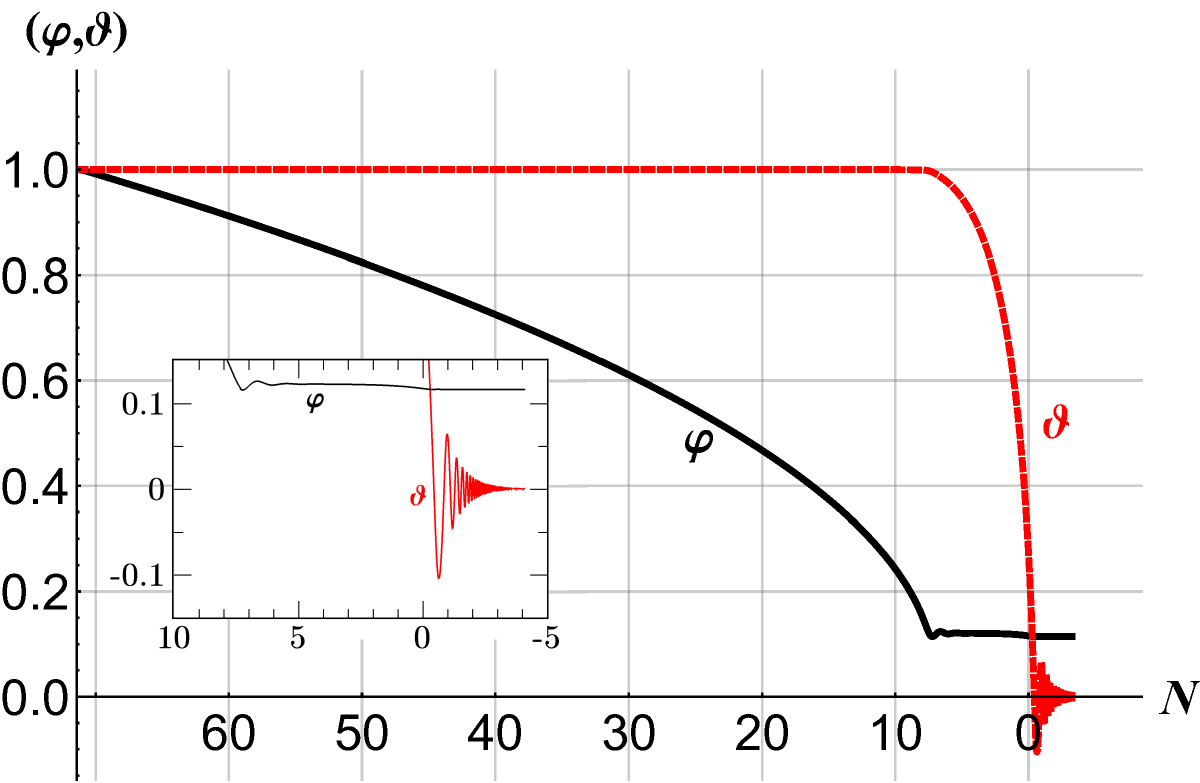}
\caption{Evolution of the Hubble parameter $H$
({\it left}) and the fields $\varphi$ (black) and $\vartheta$ (red, dashed) ({\it right}), as a function of the number of e-folds $N$ ($d N = - H \, d t$) before the end of inflation.  The end of inflation is defined by $\epsilon \equiv d\, {\rm ln} H / d N = 1$, or equivalently by $\ddot a = 0$, where $\ddot a$ is the second derivative of the scale factor with respect to cosmic time $t$.  Each field is normalized by its initial value. We set $\xi = 1$ and $M=M_{\rm pl}/10$, which fixes $v=\sqrt{99/100} \, M_{\rm pl}$, and take initial conditions $\varphi_i = 8.7
\, M_{\rm pl}, \dot{\varphi}_i = 0$, $\vartheta_i=0.97 \pi$,
$\dot{\vartheta}_i = 0$, with $\lambda = 4 \times 10^{-9}$ and $\Lambda = 2.736
\times 10^{-3}$.
The parameter dependence is illustrated in Figs.~\ref{fig:xi-plots} and
\ref{fig:Na-rH-plots}.
}
\label{fig:phichiHplot}
\end{figure*}

To simplify the analysis of the background dynamics, we consider the length of the background fields' velocity vector \cite{GrootNibbelink:2000vx,GrootNibbelink:2001qt,Seery:2005gb,Langlois:2008mn,Peterson:2010np,Gong:2011uw,Kaiser:2012ak,Gong:2016qmq,Christodoulidis:2018qdw,Christodoulidis:2019mkj,Christodoulidis:2019jsx}
\be
 \dot{\sigma} \equiv \vert \dot{\varphi}^I \vert = \sqrt{ {\cal G}_{IJ}^{(E)} \dot{\varphi}^I \dot{\varphi}^J } \, ,
\ee
where the components of the vector $\dot{\varphi}^I (t) = (\dot{\varphi} (t), \dot{\vartheta} (t) )$ consist of the derivatives (with respect to cosmic time $t$) of the spatially homogeneous background fields. We may then define a unit vector that points in the direction of the background fields' evolution:
\be
\hat{\sigma}^I \equiv \frac{\dot{\varphi}^I}{\dot{\sigma}} .
\ee
The background equations simplify to
\be
\begin{split}
H^2 &= \frac{1}{3 M_{\rm pl}^2} \left[ \frac{1}{2} \dot{\sigma}^2 + V_E \right] , \\
\ddot{\sigma}& + 3 H \dot{\sigma} + V_{E, \sigma} = 0  ,
\end{split}
\ee
where we define
\be
V_{E, \sigma} \equiv \hat{\sigma}^I V_{E, I} .
\label{Vsigma}
\ee
Thus we arrive at effectively single-field background evolution, along a direction in field space defined by $\hat{\sigma}^I$.

The evolution of the direction of the trajectory can be described by the covariant turn-rate vector,
\be
\omega^I \equiv {\cal D}_t \hat{\sigma}^I  \, ,
\label{eq:turnrate}
\ee
where ${\cal D}_t A^I \equiv \dot{\varphi}^J \, {\cal D}_J A^I$ for a vector $A^I$ in the field space, and ${\cal D}_J A^I$ is the usual covariant derivative associated with the field-space metric ${\cal G}_{IJ}^{(E)}$.  We may define the (scalar) turn rate as \cite{Achucarro:2016fby}
\begin{equation}
    \omega \equiv \epsilon_{IJ} \, \hat{\sigma}^I \, \omega^J \, ,
    \label{omegadef}
\end{equation}
where
\begin{equation}
    \epsilon_{IJ} \equiv \left[ {\rm det} \left( {\cal G}_{IJ}^{(E)} \right) \right]^{1/2} \, \bar{\epsilon}_{IJ} \> , \> \epsilon^{IJ} = \left[ {\rm det} \left( {\cal G}_{IJ}^{(E)} \right) \right]^{-1/2} \bar{\epsilon}^{IJ} \, ,
    \label{epsilondef}
\end{equation}
and $\bar{\epsilon}_{IJ} = \bar{\epsilon}^{IJ}$ is the usual Levi-Civita symbol: $\bar{\epsilon}_{12} = +1$ and $\bar{\epsilon}_{IJ}=-\bar{\epsilon}_{JI}$. Note that with this definition of the scalar turn rate, $\omega = \pm \vert \omega^I \vert$.

\begin{figure}[h!]
\includegraphics[trim=0 0in 0 0.in,clip,width=0.48\textwidth]{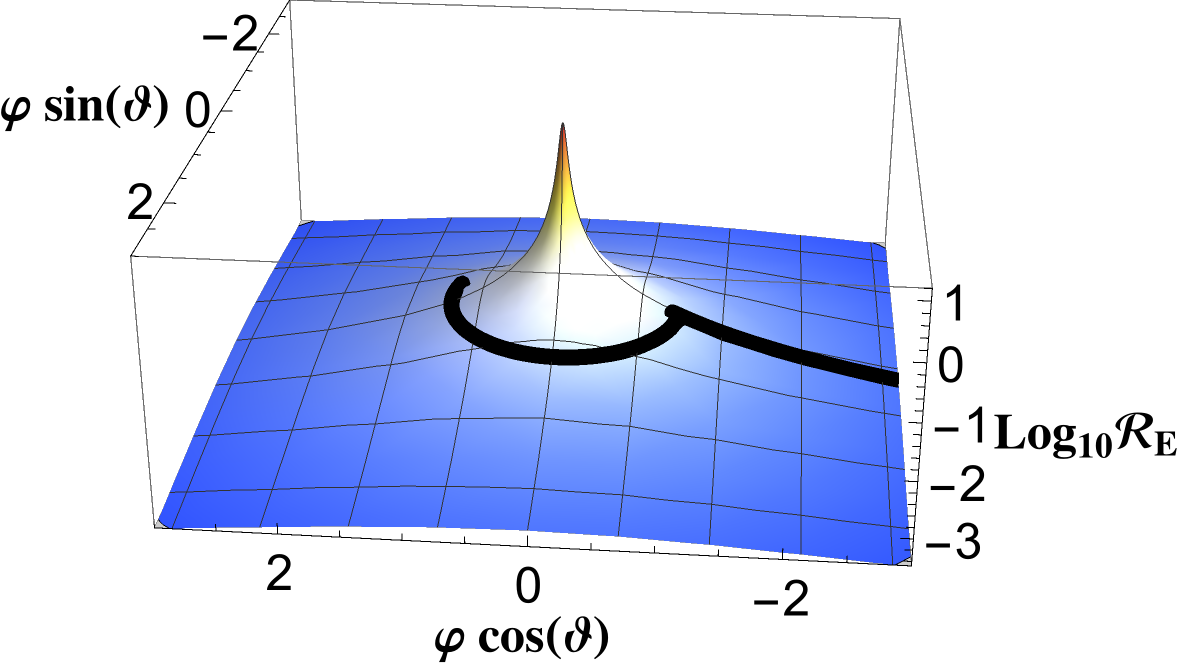}
\caption{The inflationary trajectory in field space (black line), for the example parameters of Fig.~\ref{fig:phichiHplot}, superimposed on a plot of the field-space Ricci scalar ${\cal R}_E$. We set $M_{pl}=1$. For illustrative purposes, only the end of the radial-inflation phase is shown.}
\label{fig:Plot3D-trajectory}
\end{figure}

 \begin{figure*}[t]
\centering
\includegraphics[width=0.3\textwidth]{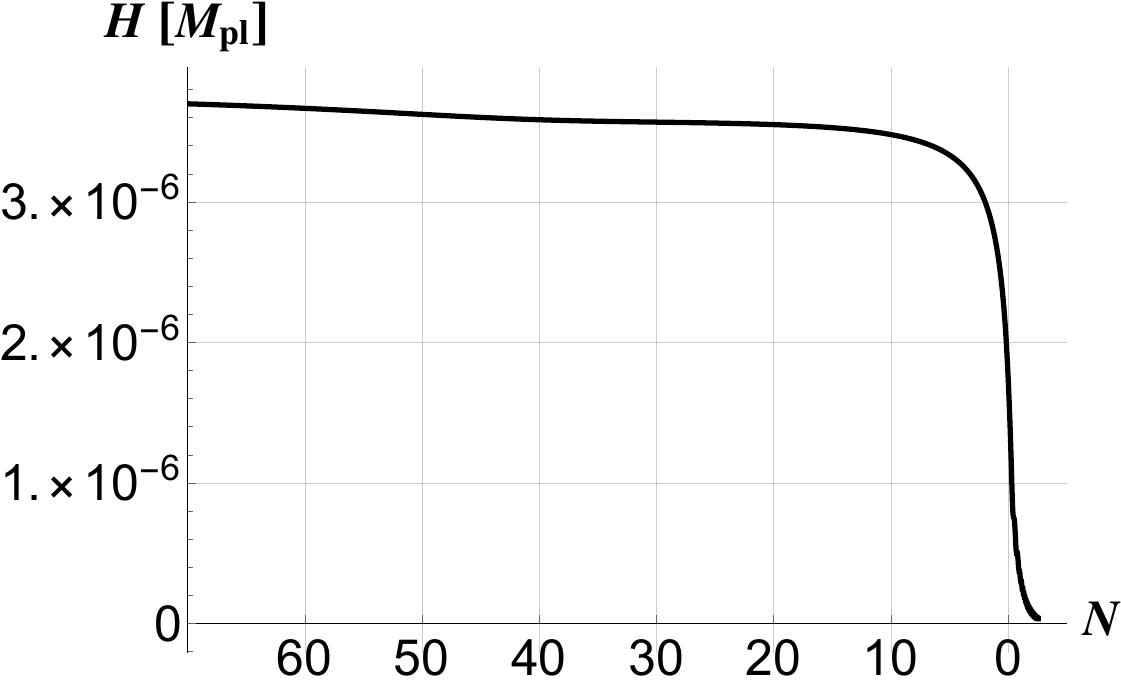}
\includegraphics[width=0.3\textwidth]{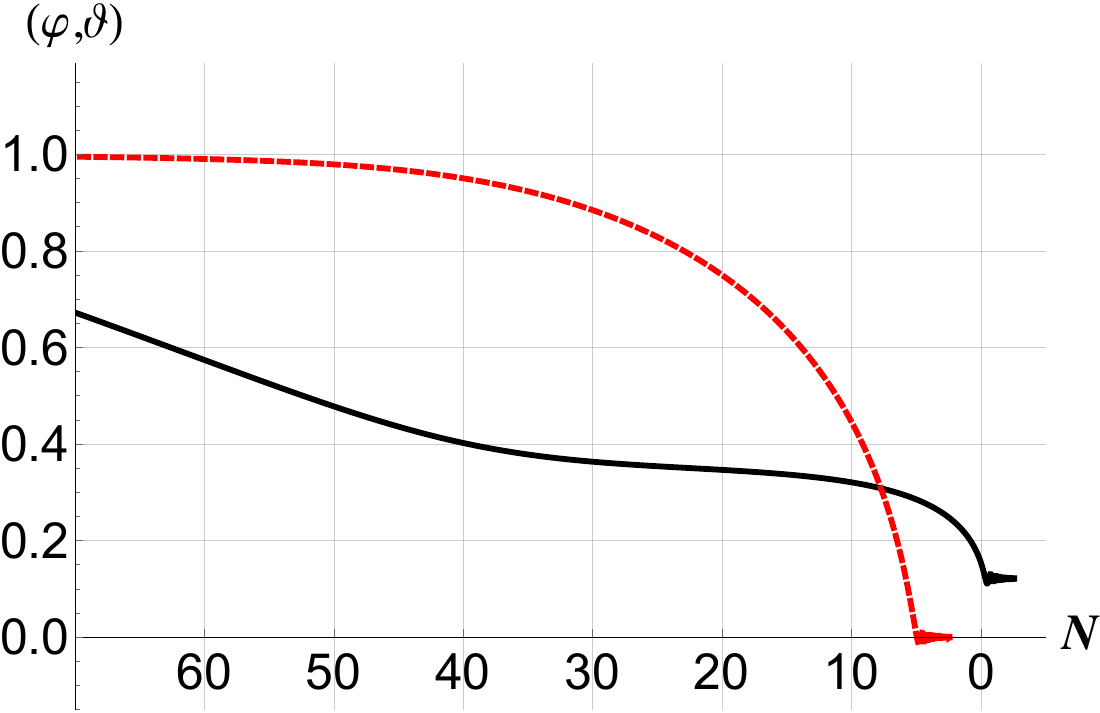}
\includegraphics[width=0.3\textwidth]{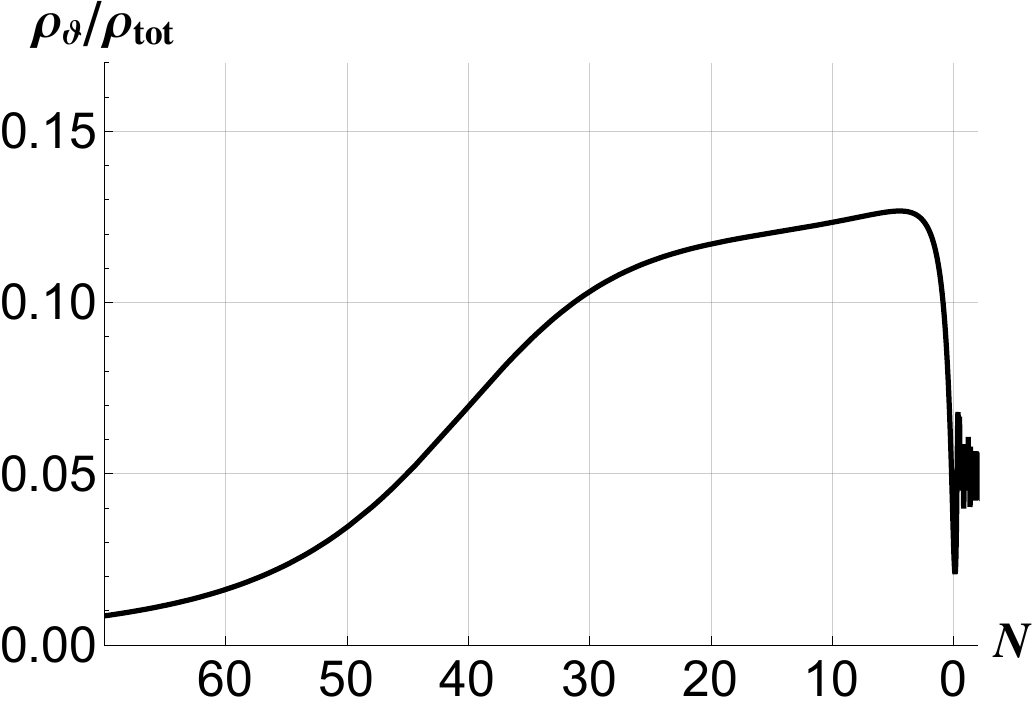}
\caption{ Evolution of the Hubble parameter ({\it left}), the
fields $\varphi$ (black) and $\vartheta$ (red, dashed) ({\it middle}), and the fraction of the energy density in the axion field ({\it right}), in a model with a very mild hierarchy between Hubble constants in the two phases.  Each field is normalized by its initial value. We set $\xi = 1$ and $v = 0.995 \, M_{\rm pl}$, and take initial conditions $\varphi_i = 8.2
\, M_{\rm pl}, \dot{\varphi}_i = 0$, $\vartheta_i=
0.952
\pi$, $\dot{\vartheta}_i = 0$, with
$\lambda = 1.739 \times 10^{-10}$ and $\Lambda = 3.562
\times 10^{-3} \,  M_{\rm pl} $.
The graphs show the final 70 e-folds of the 108.4 e-folds of inflation that result from these initial conditions.}
\label{fig:phichiHplot-2}
\end{figure*}

\section{ The Phases of Inflation
}
\label{sec:inflation}

\begin{figure*}[t]
\includegraphics[width=0.49\textwidth]{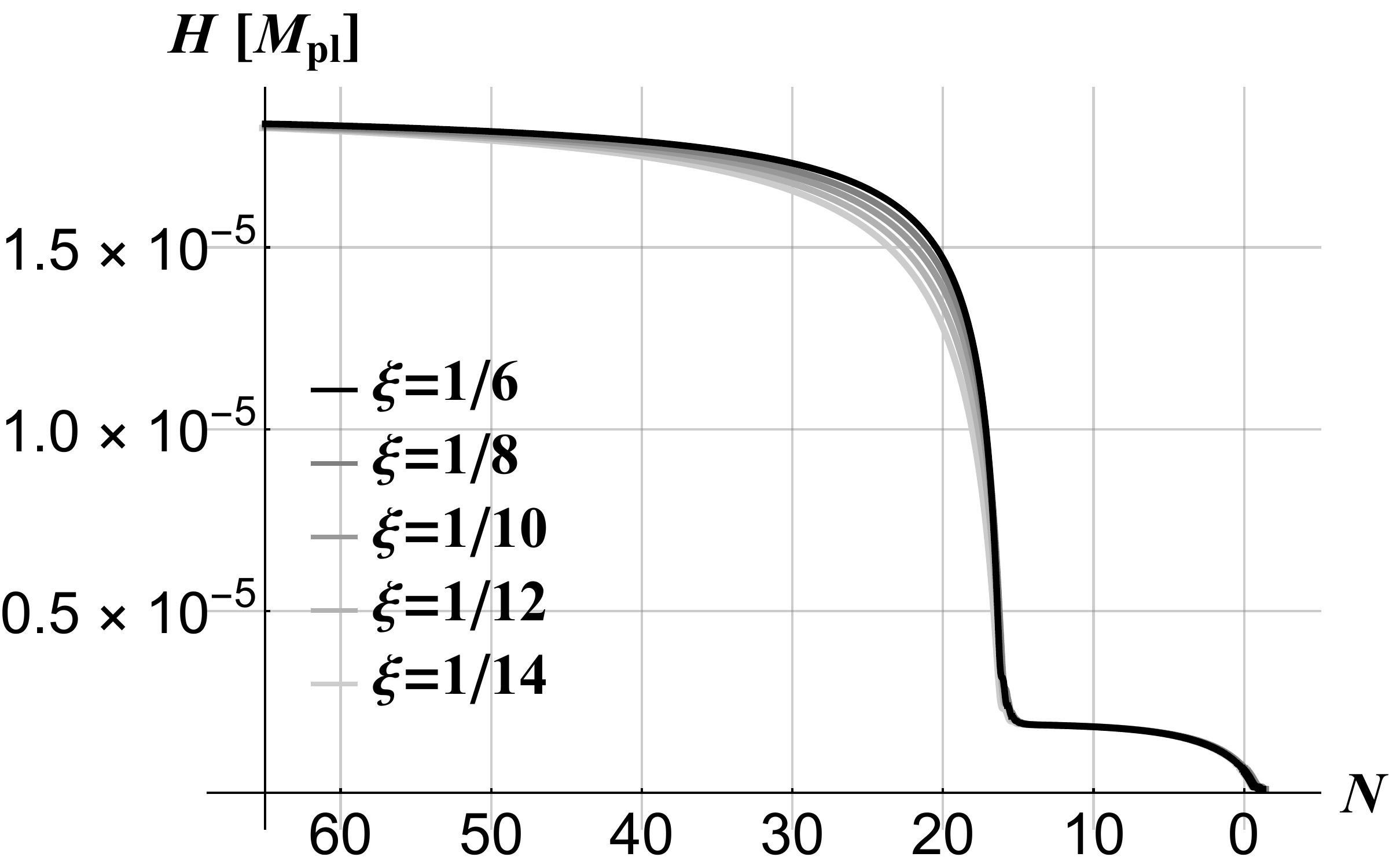}
\includegraphics[width=0.49\textwidth]{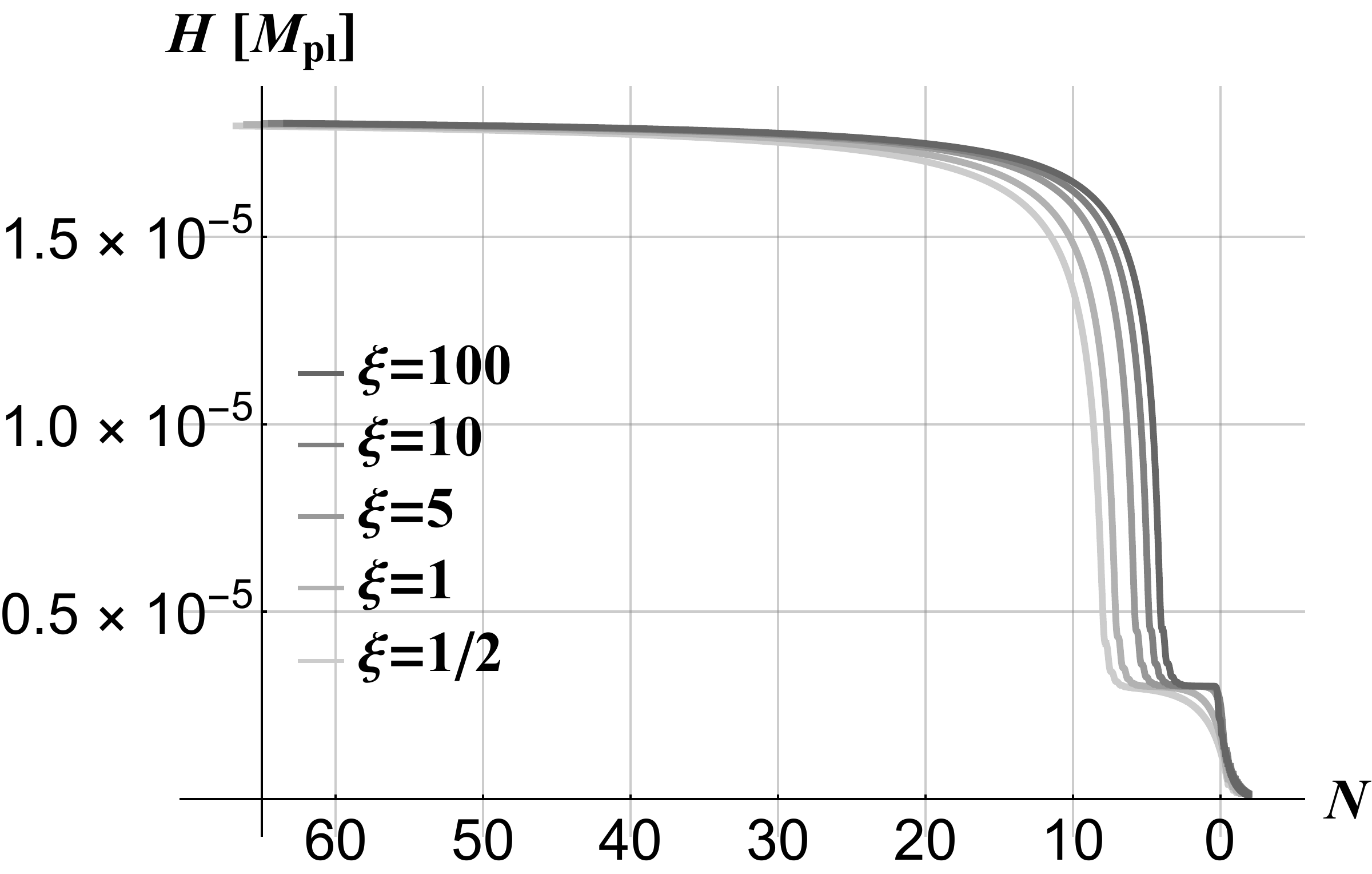}
\caption{The dynamics of radial inflation for cases with a large
hierarchy between the two inflationary scales ($\Lambda^4 \ll
\lambda v^4$). Smaller values of $\xi$ accelerate the decline of
$H$ during the radial-inflation phase, as seen in the first of Eqs.~\eqref{eq:SRparams}. {\it Left:} we fix $v=2M_{\rm pl}$ and $\Lambda = 1.529 \times 10^{-3} M_{\rm pl}$, and vary $\xi=1/6$, $1/8$, $1/10$, $1/12$, and $1/14$ (top to bottom). The initial conditions are $\vartheta_i=0.9 \pi$, $\dot{\vartheta}_i = 0$, and $\dot{\varphi}_i=0$ for all cases, and, for the purpose of visual comparison, we adjust $\lambda$ and $\varphi_i$ to keep $H_i$ and the duration of inflation approximately fixed.  {\it Right:} to allow for $\xi \gg 1$ we fix $\sqrt{\xi} v= (9/10) M_{\rm pl}$ and vary $\xi=100$, $10$, $5$, $1$, and $1/2$ (top to bottom). Initial conditions for the axion field are chosen to promote visibility by fixing the duration of axion-driven inflation (the second plateau of $H$) for each curve to be spread in the range of $5-8$ e-folds.  The $N=0$ point on the $x$-axis is taken to be the end of axion inflation (i.e., $\epsilon=1$) for each case. The full set of parameters is listed in Appendix \ref{app:params}. }
\label{fig:xi-plots}
\end{figure*}

\begin{figure}[h!]
\includegraphics[width=0.49\textwidth]{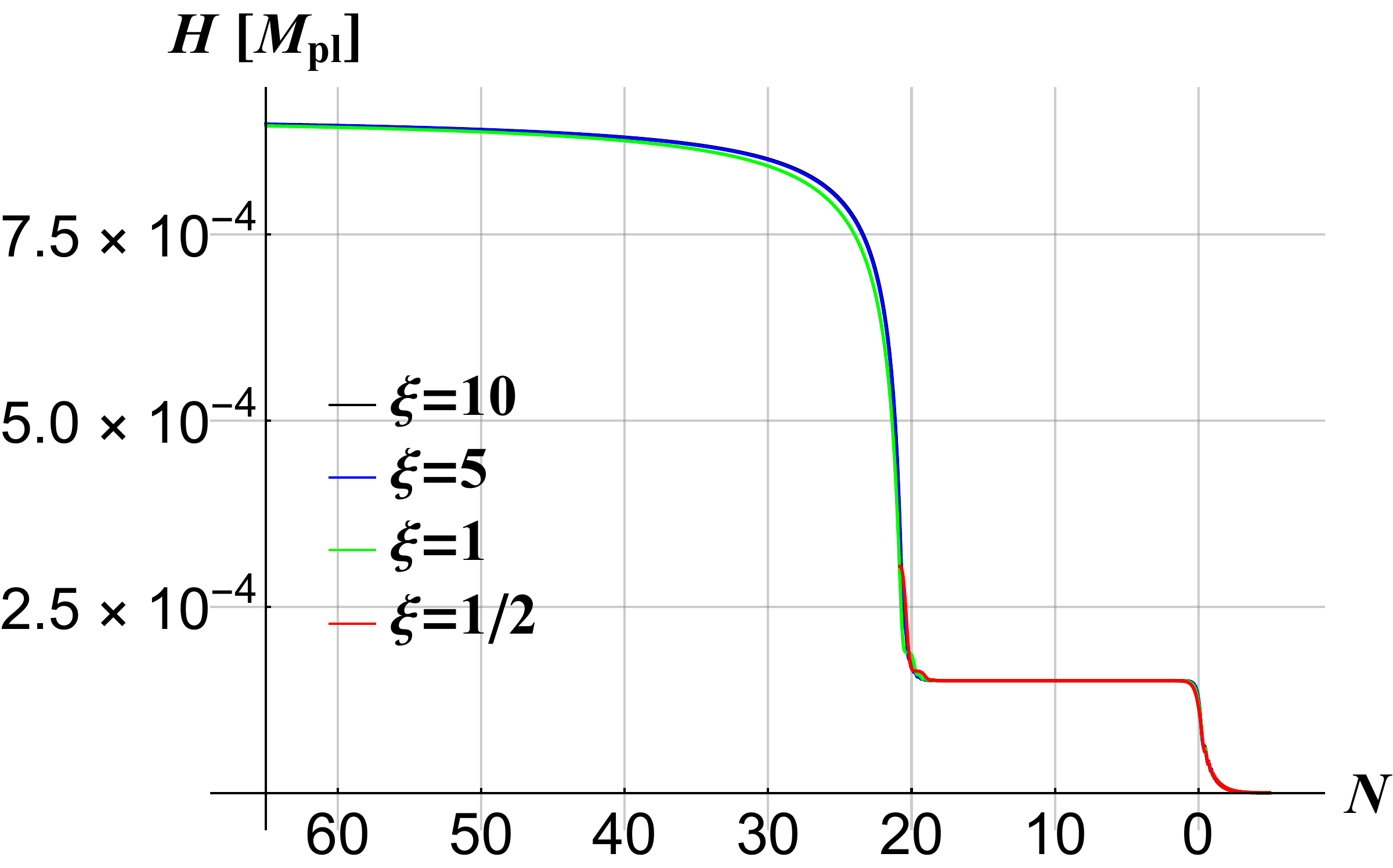}
\caption{The behavior of the Hubble parameter for a sequence of different $\xi$ parameters, with the parameter $v$ held fixed at $v=0.3 M_{\rm pl}$.  For $\xi = 10, 5,$ and 1, the diagram shows the final 65 e-folds of evolution which began somewhat earlier, while for $\xi=\frac{1}{2}$ the entire evolution is shown.The full set of parameters, which were chosen to make the four curves look almost identical, is listed in Appendix \ref{app:params}.}
\label{fig:Fig5plus}
\end{figure}

We assume initial conditions $\varphi\gg v$, $r_{\varphi} \gg 1$, and an initial displacement angle of the axion $\vartheta \sim \mathcal{O}(1)$. In this case the potential of Eq.~\eqref{eq:VE} is initially dominated by the radial field $\varphi$, and is well approximated by the plateau,
\begin{equation}
    V_{E} \sim \frac{\lambda}{4 \xi^2} M_{\rm pl}^4.
\end{equation}
As $\varphi$ decreases towards $v$ the relative contribution of the axion potential increases. This can trigger the onset of a second phase of inflation, during which the axion contributes significantly to the energy density; depending on parameters, the axion contributions can dominate the system's energy density during this second phase. 

It will be convenient to describe the background evolution of this system in terms of the number $N$ of e-folds of accelerated expansion before the end of inflation, defined as
 \begin{equation}
     {\rm d}N = - H {\rm d}t ,
 \end{equation}
where $N=0$ is taken to be the end of inflation, defined by $\epsilon \equiv {\rm d}({\rm ln} H)/{\rm d} N = 1$. The duration of inflation is then controlled both by the model parameters and the initial conditions for the fields $\varphi$ and $\vartheta$. 

A paradigmatic example is shown in Fig.~\ref{fig:phichiHplot}, which shows the evolution of the Hubble parameter and the scalar fields for $\xi = 1$ and $\vartheta_i = 0.97 \pi$. Two distinct phases of inflationary evolution are evident, with $\sim 60$ e-folds of inflation driven by the radial field $\varphi$, followed by a sudden transition to (in this case) $ \sim 10$ e-folds of inflation driven by the axion field. In Fig.~\ref{fig:Plot3D-trajectory}, this inflationary trajectory is shown in field space, superimposed on a plot of the field-space Ricci scalar ${\cal R}_E$.

In the example shown in Fig.~\ref{fig:phichiHplot} there is a clear separation between radial and axion inflation phases, and the Hubble parameter exhibits a hierarchy between the two phases. Alternatively, if $\Lambda/(\lambda^{1/4} v) = \mathcal{O}(1)$ while maintaining $\Lambda \gg v$, the second phase may be characterized by having the energy density distributed more evenly between the two fields. In this case there is no hierarchy in $H(t)$, despite a transition from a phase of inflation driven entirely by the radial field to a second phase driven by contributions from both the radial and axion fields. A paradigmatic example is shown in Fig.~\ref{fig:phichiHplot-2}, where the left panel shows the evolution of $H$, and the right panel shows the evolution of the angular-field's contribution to the total energy density, defined as,
\begin{eqnarray}
\frac{\rho_{\vartheta}}{\rho_{\rm tot}} = \frac{1}{3 M_{pl}^2 H^2}&&\left[     \frac{1}{2} \left(\frac{M_{\rm pl}^2}{2f} \right) \varphi^2  \dot{\vartheta}^2 \right. \\
&& \left.  +\frac{ M_{\rm pl}^4\Lambda^4 }{( M^2 + \xi \varphi^2)^2}\left( 1 - \cos \vartheta\right)\right]. \nonumber
\end{eqnarray}
In this example, during the latter phase of inflation about 12\% of the energy density is contributed by the axion field. 

The behavior of the model can be understood analytically. Let us first consider the radial field $\varphi$. At early times, we have $f \simeq \frac{1}{2}\xi \varphi^2$, and hence
\be 
\ddot {\varphi} + 3 H \dot{\varphi} -
\frac{\dot{\varphi}^2}{\varphi}+ \frac{\xi \varphi^2}{M_{\rm pl}^2
(1 + 6 \xi)} \, V_{E,\varphi} \simeq 0 ,
\ee
where we have neglected the term $- M^2 \dot
\vartheta^2 /[\xi(1+6 \xi)]$, because it is negligibly small in
the cases discussed in this paper. In the slow-roll limit, this simplifies further, to
\be
3 H \dot \varphi + \left( \frac{\varphi^2}{6 \alpha M_{\rm pl}^2}
\right) V_{E, \varphi} \simeq 0,
\ee
where we have defined $\alpha$ as
\be
\alpha \equiv 1 + \frac{1}{6\xi}.
\label{eq:alpha}
\ee
To make contact with the past literature (see, e.g., Refs.~\cite{Bezrukov:2007ep,Galante:2014ifa}), we define
\begin{equation}
    \phi \equiv \sqrt{ 6 \alpha} \, M_{\rm pl} \, {\rm ln} (\varphi / M_{\rm pl} ) \, .
    \label{phidef}
\end{equation}
The slow-roll equation of motion becomes simply
\be
3 H \dot{\phi} + \partial_{\phi}V_E \simeq 0.
\ee
Thus we arrive at the standard equation describing single-field slow-roll inflation of a canonically normalized scalar field.

We now turn to the potential. In the regime for which $f \simeq \frac{1}{2}\xi \varphi^2$, the potential is given by
\be
V_E \simeq \frac{\lambda}{4} M_{\rm pl}^4 \frac{ (\varphi^2 - v^2)^2}{ \xi^2 \varphi^4} + \frac{ M_{\rm pl}^4\Lambda^4 }{\xi^2 \varphi^4}\left( 1 - \cos \vartheta\right).
\label{eq:VEradial}
\ee
In terms of the rescaled radial field $\phi$, this may be written
\bea
V_E \simeq &&\frac{\lambda}{4} \frac{M_{\rm pl}^4}{\xi^2} \left(1- \frac{v^2}{M_{\rm pl}^2} e^{-\sqrt{\frac{2}{3\alpha }}\phi/M_{ \rm pl}} \right)^2 \label{VEradial}\\ &&+ \frac{\Lambda^4 }{\xi^2 } e^{-2\sqrt{\frac{2}{3\alpha }}\phi/M_{\rm pl}} \left( 1 - \cos \vartheta \right) \nonumber .
\eea
By construction, the axion term makes a subdominant contribution in this regime, since it arises from the spontaneous breaking of an approximate symmetry. The initial phase of inflation driven by the radial field is therefore well approximated as being driven by the first contribution in Eq.~(\ref{VEradial}), which we may write as
\be
\label{eq:Vesimple}
V_E \simeq V_0 \left(1- \frac{v^2}{M_{\rm pl}^2}e^{-\sqrt{\frac{2}{3\alpha}}\phi/M_{\rm pl}} \right)^2 ,
\ee
where we define $V_0 \equiv (\lambda M_{\rm pl}^4 )/( 4\xi^2)$. This form of the potential is familiar as a class of inflation models known as $\alpha$-attractors \cite{Ferrara:2013rsa,Kallosh:2013yoa,Kallosh:2014rga}. 

\begin{figure*}[t]
\includegraphics[width=0.49\textwidth]{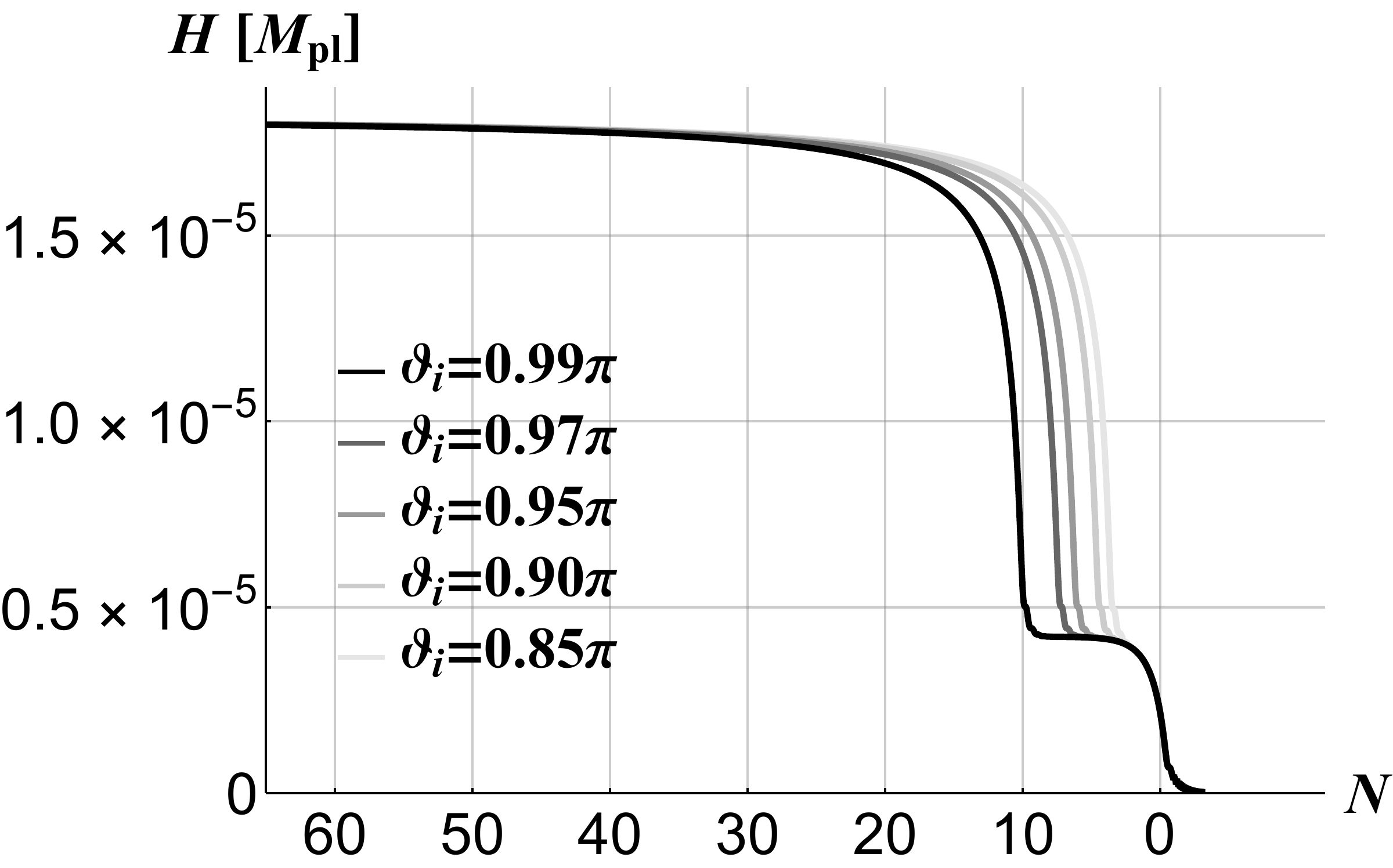}
\includegraphics[width=0.49\textwidth]{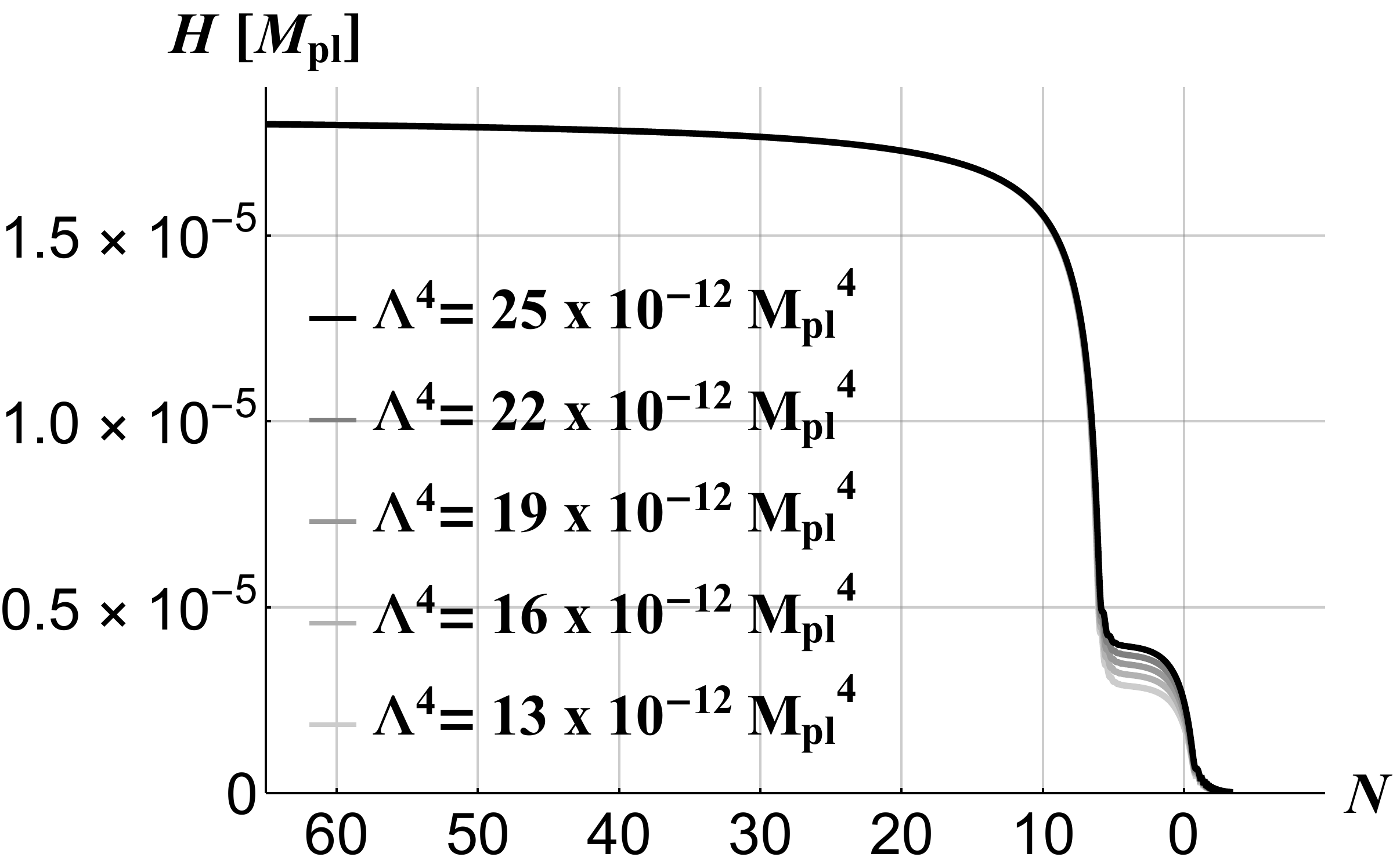}
\caption{ The dynamics of axion inflation for cases with a large
hierarchy between the two inflationary scales ($\Lambda^4 \ll
\lambda \, v^4$). 
{\it Left:} the number of e-folds of axion inflation, $N_a$, is set by the initial condition for the axion field. We consider the example of Fig.~\ref{fig:phichiHplot}, but with differing initial conditions, $\vartheta_i/\pi=0.85$, $0.9$, $0.95$, $0.975$, and $0.99$ (left to right). The $N=0$ point on the $x$-axis is the end of inflation in the $\vartheta_i/\pi=0.85$ example.  {\it Right:} the energy scale of the axion-inflation phase is set by $\Lambda$. We fix $\vartheta_i=0.95 \pi$ and consider differing values $\Lambda^4/(10^{-12} M_{\rm pl}^4) = 25$, $22$, $19$, $16$, and $13$ (top to bottom). The full set of parameters are listed in Appendix \ref{app:params}. }
\label{fig:Na-rH-plots}
\end{figure*}

During the slow-roll phase of radial inflation, when $v^2 e^{-\sqrt{\frac{2}{3\alpha}}\phi/M_{\rm pl}} \ll M_{\rm pl}^2$, the radial field $\varphi$ evolves as
\be
\varphi(N) \simeq v\sqrt{\frac{4 N }{3 \alpha}} 
\label{eq:phiN}
\ee
for $N\gg1$, where $N$ measures the number of e-folds before the end of inflation, $dN = - H dt$.  The evolution of the Hubble parameter is described by the (Hubble) slow-roll parameters, $\epsilon \equiv {\rm d} ({\rm ln} H) / {\rm d} N$ and $\eta \equiv 2\epsilon + ( 2 \epsilon)^{-1} ({\rm d}\epsilon/{{\rm d}N})$, which, for $\alpha \ll N$, take the form
\be
\label{eq:SRparams}
\epsilon(N)= \frac{3 \alpha}{4 N^2} \;\; , \;\; \eta (N) =\frac{1}{N}.
\ee
If we were considering a single-field model with the potential in Eq.~\eqref{eq:Vesimple}, then the inflationary predictions for the spectral index $n_s$ and tensor-to-scalar ratio $r$ would follow from these expressions for $\epsilon$ and $\eta$, and be given by
\be
n_s = 1 - \frac{2}{N_r} \;\; , \;\; r = \frac{12 \alpha}{N_r ^2} \;\; ,
\label{eq:radialnsr}
\ee
where $N_r$ is the number of e-folds before the end of radial inflation when the CMB pivot scale exited the horizon. These predictions match those of standard $\alpha$-attractor models \cite{Kallosh:2013yoa,Kallosh:2015lwa,Christodoulidis:2018qdw,Christodoulidis:2019jsx}. In the limit of $\xi \gg 1$, or equivalently $\alpha \rightarrow 1$, the predictions match those of the Starobinsky model \cite{Starobinsky:1980te}, Higgs inflation \cite{Bezrukov:2007ep}, and, more generally, the attractor behavior of multifield models with nonminimal couplings in the limit $\xi_I \gg 1$ \cite{Kaiser:2013sna}.  However, as we will see in Sec.~\ref{sec:observables}, these predictions can be considerably modified in our two-field model, in certain regions of parameter space.

Recalling Eq.~(\ref{eq:alpha}), which relates $\alpha$ to $\xi$, we see that the evolution of $\varphi$ encodes the $\xi$-dependence of the model. The symmetry-breaking scale $v$ enters the dynamics as a multiplicative shift of $\varphi$, as in Eq.~(\ref{eq:phiN}), while $\lambda$ enters only via the height of the plateau $V_0$. Thus the dynamics of the radial phase of inflation are determined by $\xi$. This can be appreciated from Fig.~\ref{fig:xi-plots}, in which we consider the evolution of the Hubble parameter for differing values of $\xi$. In the left panel we fix $v=2 M_{\rm pl}$ and the axion initial conditon $\vartheta_i =0.9 \pi$, and (as usual) adjust $M$ such that $\xi v^2 + M^2 = M_{\rm pl}^2$, which (for $v = 2 M_{\rm pl})$ requires $\xi < 1/4$. In the right panel, we instead fix the combination $\sqrt{\xi} \, v = 0.9 \, M_{\rm pl}$, which allows us to consider a larger range of values for $\xi$.

From the appearance of Fig.~\ref{fig:xi-plots}, one might conclude that the ending of the axion phase of inflation is strongly influenced by $\xi$, with the ending being more gently rounded for lower values of $\xi$, and sharper for higher values of $\xi$.  This feature, however, is not caused by the variation of $\xi$.  One can see from Eq.~(\ref{eomaxion}) that if $\varphi$ is fixed at $v$, so $f = \frac{1}{2} (M^2 + \xi v^2) = \frac{1}{2} M_{\rm pl}^2$, then the equations of motion for $\vartheta$ are independent of $\xi$.  They are strongly influenced by $v$, however, since ${\cal G}^{\vartheta\vartheta} \propto 1/\varphi^2 \approx 1/v^2$.  The parameters in this figure were chosen so that $\sqrt{\xi} \, v$ was fixed, so larger $\xi$ implies smaller $v$, which in turn implies a strong force acting on the axion field $\vartheta$.  The effect of varying $\xi$ in this range of larger $\xi$ values, while holding $v$ fixed, is shown in Fig.~\ref{fig:Fig5plus}, which shows that there is no visible effect.  The effect of varying $\xi$ over a range of smaller values, while holding $v$ fixed, is shown in Fig.~\ref{fig:xi-plots}, where one also sees that the evolution of axion inflation is uninfluenced by $\xi$.

Now we turn to a more detailed discussion of the axion field. During the phase of radial inflation, the axion is effectively massless, due to the exponential suppression of the axion potential. In particular, the axion mass reduces to $m_{\vartheta}^2 \simeq {\cal G}^{\vartheta\vartheta} V_{E, \vartheta \vartheta}$. Compared to the Hubble scale during radial inflation, we find
\begin{equation}
    \frac{\vert m_{\vartheta}^2 \vert }{H^2} \simeq 
\frac{3 \Lambda^4}{\xi V_0}
    e^{-2 \sqrt{ \frac{ 2}{3\alpha} } \, \phi / M_{\rm pl} } \ll 1 \, .
    \label{eq:mchiH}
\end{equation}
For $\Lambda^4 < V_0$, the axion remains ultra-light during radial inflation. In that regime, with $\xi \varphi^2 \gg M^2$, we also find $\Gamma^\theta_{\vartheta \varphi} \dot \varphi
\ll H$,
and hence Eq.~(\ref{eomaxion}) reduces to
\be
\ddot{\vartheta} + 3H\dot\vartheta  \simeq 0.
\ee
During radial inflation, the axion obeys $|\dot \vartheta| \ll H$, and remains effectively frozen in place.

As $\varphi$ decreases, the energy density in the axion increases, leading to a second phase of inflation characterized by a non-negligible energy density in the axion. In the simple case that the axion comes to dominate, as in Fig.~\ref{fig:phichiHplot}, the second phase is well characterized by conventional natural inflation. The number of e-folds of axion inflation is given by \cite{Martin:2013tda}
\be
\label{Naxion}
N_a \simeq - \frac{v^2}{M_{\rm pl}^2} {\rm ln}\, \left[ \frac{M_{\rm pl}^2 + 2 v^2}{4 v^2}\left( 1+ \cos \vartheta_i\right) \right],
\ee
where $\vartheta_i$ is the value of $\vartheta$ at the onset of axion inflation. Since $\vartheta$ undergoes ultra-slow-roll evolution during the radial-inflation phase, the value of $\vartheta$ at the start of the axion-inflation phase is virtually unchanged from its value at the onset of radial inflation. The dependence of the duration of this second phase of inflation, $N_a$, on the axion initial condition, $\vartheta_i$, is shown in the left panel of Fig.~\ref{fig:Na-rH-plots}. During this phase of natural inflation, the Hubble parameter is given by
\be
H_a^2 \simeq \frac{2 \Lambda^4}{3 M_{\rm pl}^2} .
\ee
In the right panel of Fig.~\ref{fig:Na-rH-plots} we show the evolution of $H$ for varying values of $\Lambda$. 

\section{Perturbations}
\label{sec:perts}

We now proceed to study the dynamics of perturbations. We will do so both analytically and numerically. For concreteness we specialize to four fiducial model examples, which together exemplify the possibilities of the model. The parameters are chosen to each produce a spectral index $n_s=0.965$ and amplitude of the primordial (adiabatic) power spectrum $A_s=2.1\times10^{-9}$ (for a CMB pivot scale that exited the horizon $N_*=55$ e-folds before the end of inflation), despite differing parameters and cosmological dynamics. We leave a detailed exploration of parameter space to future work. The evolution of $H$ for each example is shown in Fig.~\ref{fig:H-plot-perts}, and the parameters for each are given in Appendix \ref{app:params}. 

 \begin{figure}[h!]
\includegraphics[width=0.48\textwidth]{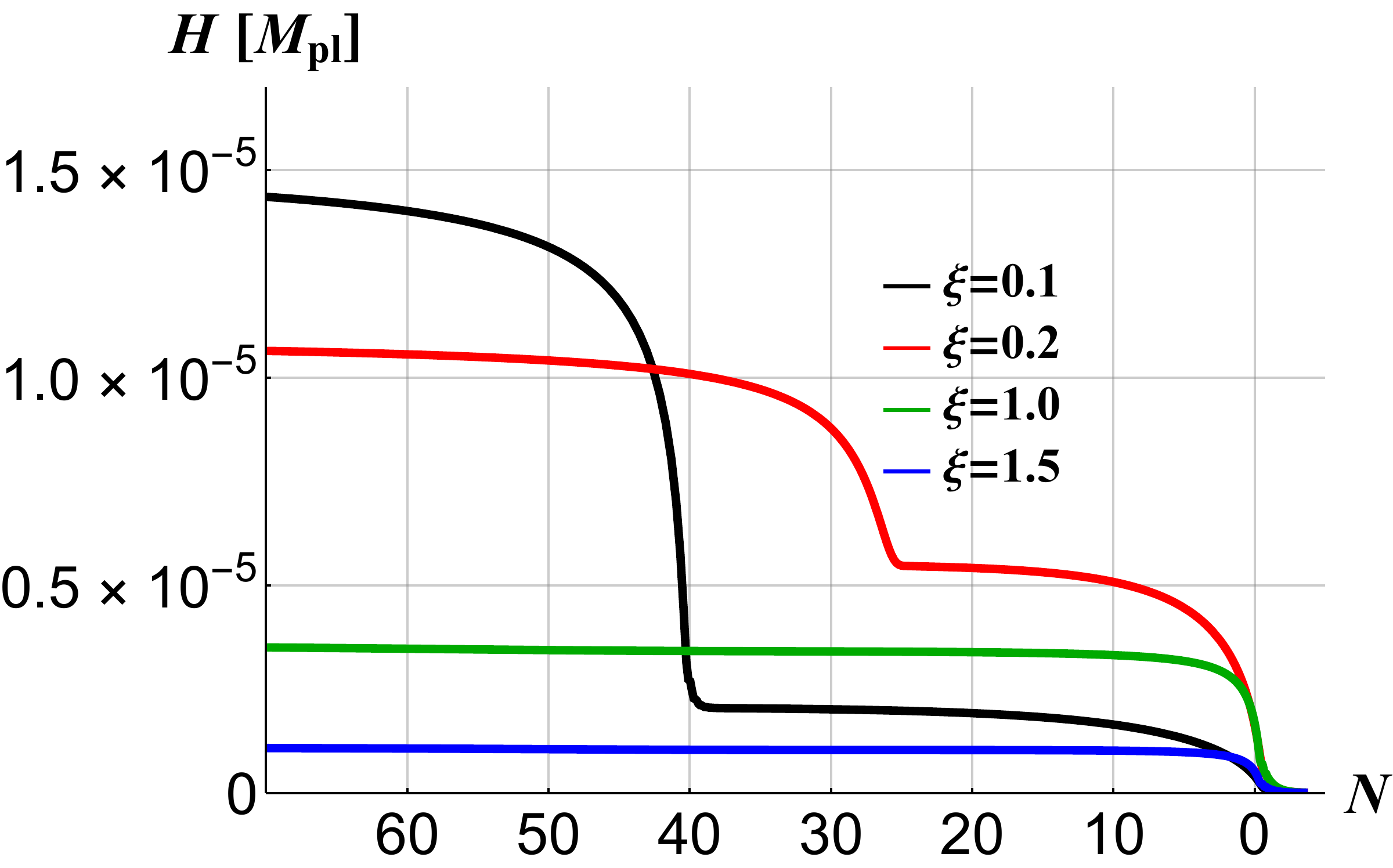}
\caption{Evolution of the Hubble parameter in the four fiducial examples to be studied numerically. Parameters for each example are given in Appendix \ref{app:params}. The red and black curves lie in the class of examples exemplified by Fig.~\ref{fig:phichiHplot}, whereas the green and blue curves lie in the class of example exemplified by Fig.~\ref{fig:phichiHplot-2}.}
\label{fig:H-plot-perts}
\end{figure}

 \begin{figure}[h!]
\includegraphics[trim={0cm 0cm 0cm 0cm},clip,width=0.49\textwidth]{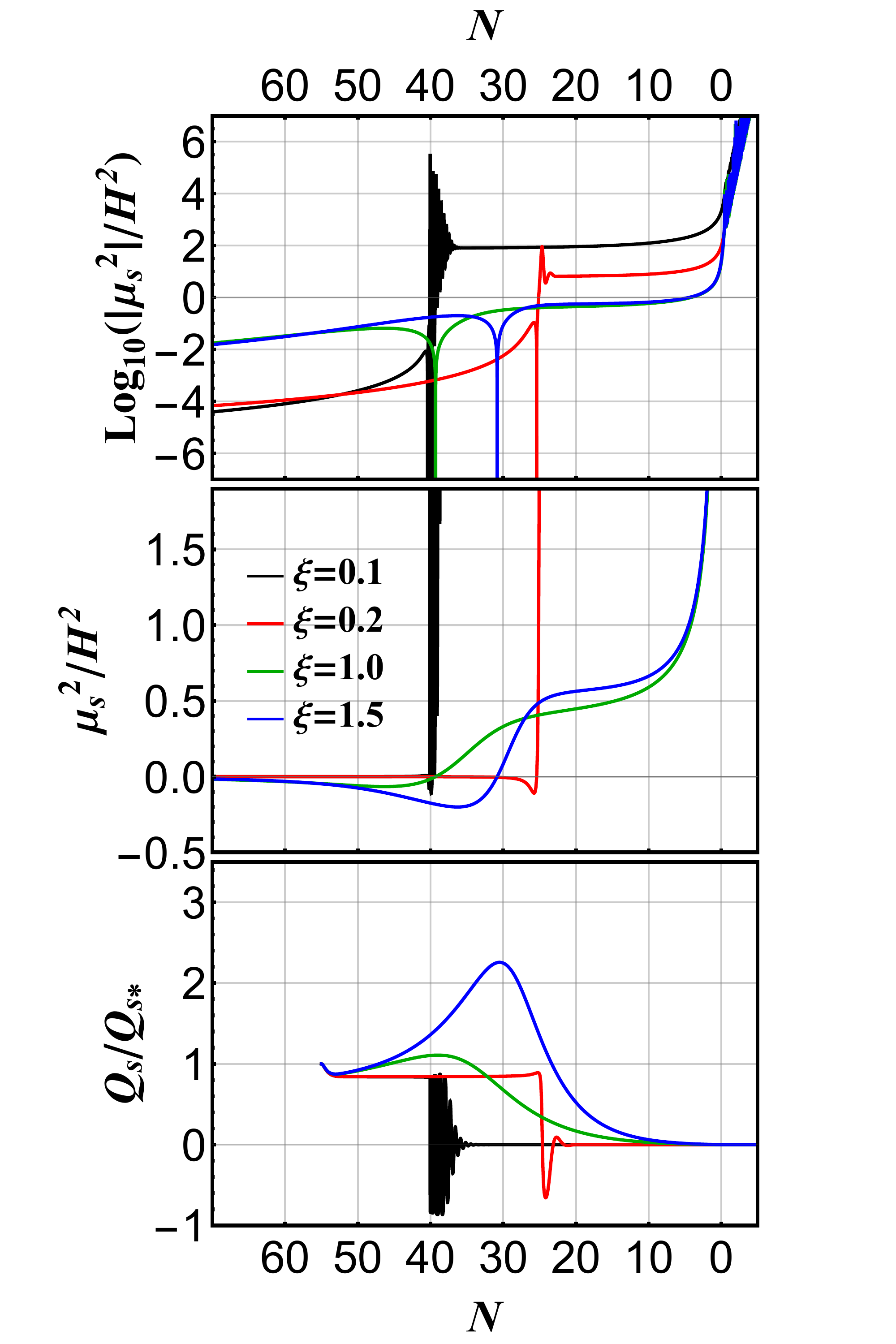}
\caption{  Evolution of Isocurvature Perturbations. ({\it Top}) The ratio of the mass of the isocurvature perturbations to the Hubble parameter, $\vert \mu_s \vert / H$, for  $\xi=0.1$ (black), $\xi=0.2$ (red), $\xi=1.0$ (green), and $\xi=1.5$ (blue). The mass undergoes a sharp transition from $\vert \mu_s \vert \ll H$ to $\mu_s \gg H$ at the transition from radial to axion-driven inflation.  ({\it Middle}) The mass of the isocurvature perturbations  becomes maximally tachyonic at the transition from radial to axion inflation, as the radial field rapidly approaches its minimum value, $\varphi \simeq v$.  ({\it Bottom}) The brief period with $\mu_s^2 < 0$ and $\vert \mu_s^2 \vert / H^2 > 1$ amplifies modes $Q_s (k, t)$ on long length-scales, $k \ll aH$, compared to the modes' magnitude following Hubble crossing, $Q_{s*}$. The mode shown here has comoving wavenumber $k_*$ corresponding to the CMB pivot scale, which we take to be the mode $Q_s (k_*,t)$ that crossed outside the Hubble radius 55 e-folds before the end of inflation. { For each plot, the curves extend}
to the end of inflation, when $\epsilon(N_{\rm end})=1$, { which is equivalent to $\ddot{a} (t_{\rm end}) = 0$}. Initial conditions and other parameters for the curves shown here are listed in Appendix \ref{app:params}.}
\label{fig:mus-Qs-combined}
\end{figure}

We expand each field about a homogeneous background value,
\be
\phi^I (x^\mu) = \varphi^I (t) + \delta \phi^I (x^\mu) ,
\label{phivarphi}
\ee
with $I=1,2$ corresponding to $\varphi$ and $\vartheta$, respectively. We construct the gauge-invariant Mukhanov-Sasaki variables for the perturbations, which to first order in perturbations read \cite{Gong:2016qmq}
\be
Q^I \equiv \delta \phi^I + \frac{\dot{\varphi}^I}{H} \psi \, ,
\label{eq:MSvar}
\ee
where $\psi (x^\mu)$ is the scalar metric perturbation on comoving spatial sections.  We may project the vector $Q^I$ into components parallel and perpendicular to the background fields' motion \cite{Gordon:2000hv},
\begin{equation}
    Q^I = \hat{\sigma}^I \, Q_\sigma + \epsilon^{JI} \, \hat{\sigma}_J \, Q_s \, ,
    \label{QsigmaQs}
\end{equation}
where (for our two-field model) the adiabatic ($Q_\sigma$) and isocurvature ($Q_s$) perturbations are each scalar quantities. The masses of the perturbations are given by the mass-squared matrix \cite{Kaiser:2012ak}
\be
\label{eq:mass-squared}
{\cal M}^I \,_J \equiv {\cal G}_{(E)}^{IK} {\cal D }_J {\cal D}_K V_E - {\cal R}^{I} \,_{LMJ} \dot{\varphi}^L \dot{\varphi}^M \, ,
\ee
where ${\cal R}^I_{\>\> LMJ}$ is the Riemann tensor associated with the field-space metric ${\cal G}_{IJ}^{(E)}$. We may then identify the canonically normalized comoving curvature perturbation and isocurvature perturbation as \cite{Gordon:2000hv,Kaiser:2012ak}
\be
{\cal R} = \frac{H}{\dot{\sigma}} Q_\sigma ,
\ee
and
\be
\label{eq:Sdef}
{\cal S} = \frac{H}{\dot{\sigma}} Q_s.
\ee
The equation of motion for $Q_\sigma$ is given by \cite{Kaiser:2012ak}
\bea
\ddot{Q}_\sigma && + 3 H \dot{Q}_\sigma + \left[ \frac{k^2}{a^2} 
+ {\cal M}_{\sigma\sigma} - \omega^2 - \frac{1}{M_{\rm pl}^2 a^3} \frac{d}{dt} \left( \frac{a^3 \dot{\sigma}^2}{H} \right) \right] Q_\sigma \nonumber \\ &&= 2 \frac{d}{dt} \left( \omega Q_s \right) - 2 \left( \frac{ V_{, \sigma}}{\dot{\sigma}} + \frac{\dot{H}}{H} \right) \left( \omega Q_s \right) ,
\label{eomQsigma}
\eea
where $\omega$ is the covariant turn rate defined in Eq.~\eqref{omegadef}, and ${\cal M}_{\sigma\sigma}$ is given by
\be
{\cal M}_{\sigma\sigma} \equiv \hat{\sigma}_I \hat{\sigma}^J {\cal M}^I \,_J = \hat{\sigma}^I \hat{\sigma}^J {\cal D }_I {\cal D}_J V_E \, .
\ee
Note that the symmetry properties of the Riemann curvature tensor prevent the curvature term in Eq.~(\ref{eq:mass-squared}) from contributing to ${\cal M}_{\sigma\sigma}$, since all four indices are contracted with $\hat{\sigma}^I$ \cite{Kaiser:2012ak}; this simplification is unique to the adiabatic perturbations and does not occur for the isocurvature perturbations.

Equation (\ref{eomQsigma}) becomes more transparent when written in terms of ${\cal R}$ and ${\cal S}$. For a two-field model, this reads (see, e.g., Ref.~\cite{Achucarro:2016fby})
\be
\label{eq:zetaEOM}\frac{d}{dt}  \left(\dot {\cal R} - 2 \omega {\cal S} \right) \\
+ (3 + \delta) H \left( \dot {\cal R} - 2 \omega {\cal S} \right) + \frac{k^2}{a^2}{\cal R} = 0 \, ,
\ee
where $\delta \equiv \dot{\epsilon} / (H \epsilon) = 4 \epsilon - 2 \eta$. One can appreciate that ${\cal R}$ is massless, consistent with the conservation of the gauge-invariant curvature perturbation on super-Hubble length-scales in the absence of isocurvature perturbations \cite{Gordon:2000hv,Wands:2000dp,Senatore:2012ya}. Indeed, on large scales (making no assumptions about slow-roll or slow-turn evolution of the background fields) one finds the familiar solution,
\be
\label{eq:dotRS}
\dot{\cal R} = 2 \omega {\cal S}  \, ,
\ee
indicating that the curvature perturbation ${\cal R}$ will only evolve (on super-horizon scales) if the background evolution of the system includes a turn, with $\omega \neq 0$.

On the other hand, the isocurvature perturbations are in general massive. The equation of motion for $Q_s$ in a two-field model is given by \cite{Kaiser:2012ak,Kaiser:2013sna,Schutz:2013fua},
\be
\begin{split}
\ddot{Q}_s + 3 H \dot{Q}_s &+ \left[ \frac{k^2}{a^2} + {\cal M}_{ss} + 3 \omega^2  \right] Q_s = 4 M_{\rm pl}^2 \frac{\omega}{\dot{\sigma}} \frac{k^2}{a^2} \psi ,
\end{split}
\label{eomQsa}
\ee
where
\begin{equation}
    {\cal M}_{ss} \equiv \hat{s}^{I J} \, {\cal M}_{I J} 
    \label{Mssdef}
\end{equation}
and $\hat{s}^{IJ} \equiv {\cal G}^{IJ}_{(E)} - \hat{\sigma}^I \hat{\sigma}^J$. { (One may use the $00$ and $0i$ components of the Einstein field equations to relate $\psi$ to $\dot{\cal R}$ and ${\cal S}$ \cite{Kaiser:2012ak}; the form of Eq.~(\ref{eomQsa}) is convenient for understanding the behavior of modes $Q_s (k, t)$ in the limit $k \ll aH$.)} On large scales (again without requiring slow-roll), Eq.~(\ref{eomQsa}) reduces to
\be
\ddot{Q}_s + 3 H \dot{Q}_s + \mu_s ^2 Q_s \simeq 0 ,
\label{eomQs}
\ee
where $\mu_s$ is the effective mass of the isocurvature perturbations, given by
\be
\mu_s^2 \equiv {\cal M}_{ss} + 3 \omega^2 .
\label{mus}
\ee
From this it follows that massive isocurvature perturbations decay as $Q_s \propto a^{-3/2}$, whereas massless isocurvature perturbations effectively freeze-out on super-horizon scales, analogous to the curvature perturbation. 

\begin{figure}[h]
\includegraphics[width=0.48\textwidth]{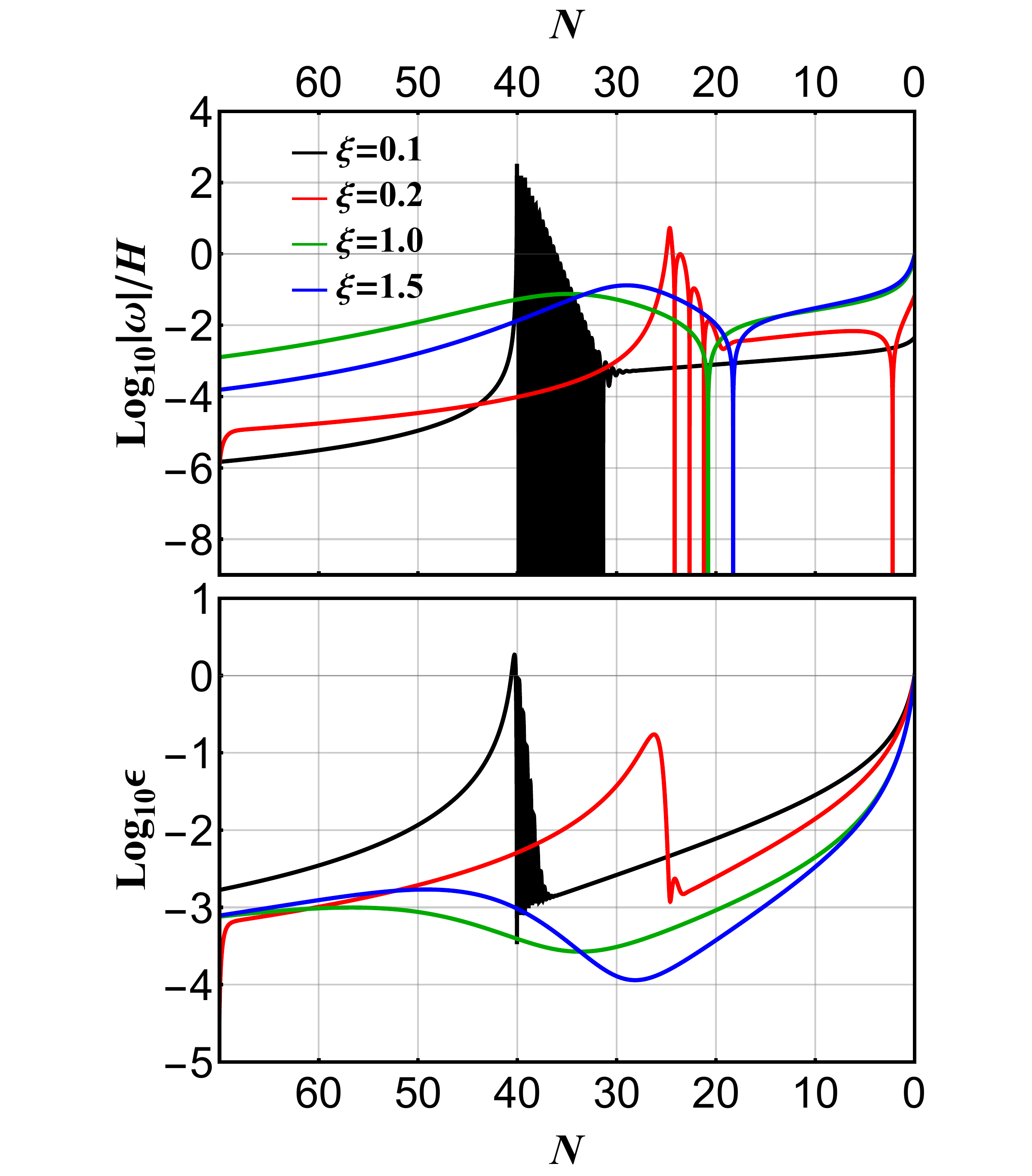}
\caption{ Background Quantities. ({\it Top}) The ratio of the covariant turn rate to the Hubble parameter, $\vert \omega \vert / H$, for  $\xi=0.1$ (black), $\xi = 0.2$ (red), $\xi= 1.0$ (green), and $\xi= 1.5$ (blue). The turn rate exhibits a dominant spike and a series of subdominant peaks at the time of the turn from radial to axion inflation. 
In each case, after the dominant spike the turn rate oscillates through zero, so $\log(|\omega|/H)$ diverges to $-\infty$. ({\it Bottom}) The slow-roll parameter $\epsilon = - \dot{H} / H^2$ for $\xi = 0.1$ (black), $\xi = 0.2$ (red), $\xi = 1.0$ (green), and $\xi = 1.5$ (blue). For each value of $\xi$, the peak of ${\cal S} (k, t)$ around the time of the turn from radial to axion inflation, shown in Fig.~\ref{fig:SRplot}, corresponds to the first {\it trough} in $\epsilon$. Initial conditions and other parameters for the curves shown here are listed in Appendix \ref{app:params}. { In both plots, each curve extends to the end of axion inflation, when $\epsilon(N_{\rm end})=1$. Initial conditions and other parameters for the curves shown here are listed in Appendix \ref{app:params}.} }
\label{fig:omegaepsplot}
\end{figure}

The behavior of $\mu_s^2$ for our two-field model is shown in Fig.~\ref{fig:mus-Qs-combined}, for $\xi$ in the range $1/10 \leq \xi \leq 1.5$. (Full sets of parameters relevant to the plot are listed in Appendix B.) We note that $|\mu_s|/H \ll 1$ at early times, whereas $\mu_s /H \gg 1$ at late times. Each of these examples exhibits a turn from purely radial inflation to either axion or mixed radial-axion inflation, as indicated by the evolution of the scalar turn rate $\omega$ of Eq.~(\ref{omegadef}), shown in Fig.~\ref{fig:omegaepsplot}. 

The evolution of the isocurvature perturbation $Q_s$ is sensitive to this behavior of $\mu_s$. We numerically solve Eq.~\eqref{eomQsa} for modes $Q_s (k, t)$ that exit the Hubble radius early during the radial-inflation phase.  We are interested in the behavior of modes after they have crossed outside the Hubble radius, so we neglect the source term on the right-hand side of Eq.~(\ref{eomQsa}), which is suppressed for $k
\ll aH$. The amplitude of modes $Q_s (k, t)$ freezes soon after
Hubble crossing, with amplitude $Q_{s*}(k)$.  Assuming that the modes begin in the Bunch-Davies vacuum state, and adopting the normalization conventions of Ref.~\cite{Gordon:2000hv}, we may approximate the amplitude of these modes as
\begin{equation}
    Q_{s*} (k) \simeq \frac{H_*}{ \sqrt{2 k^3} } \, ,
\end{equation}
consistent with a (nearly) scale-invariant power spectrum for an effectively massless scalar field.\footnote{In more detail, the random variable $Q_{s*}(k)$ is approximated by $\frac{H_*}{\sqrt{2 k^3} } e(k)$, where $e(k)$ is a Gaussian random variable satisfying $\langle e(k) \rangle = 0$ and $\langle e(k) e^{*}(k') \rangle = \delta^{3}(k-k')$, where angle brackets denote the ensemble average.} At late times, after the turn, when $\mu_s \gg H$, modes $Q_s (k, t) $ undergo damped oscillations.  At the interface between these regimes, around the time of the turn in field space, long-wavelength modes $Q_s (k, t)$ undergo a brief period of tachyonic amplification. The mass $\mu_s$ becomes maximally tachyonic immediately before the turn, as the amplitude of the radial field rapidly decreases, consistent with Eq.~(\ref{eq:mchiH}). See Fig.~\ref{fig:mus-Qs-combined}. 

The amount by which modes $Q_s (k, t)$ are amplified around the time of the turn from radial to axion inflation increases with $\xi$ and with the initial value of the axion field, $\vartheta_i$. This can be understood from the evolution of the effective mass $\mu_s ^2$, which is tachyonic before the turn, and which becomes maximally tachyonic around the time of the turn, when $\varphi \simeq v$. The numerical solution for $Q_s$ for the mode $k=k_*$ corresponding to the CMB pivot scale is shown in Fig.~\ref{fig:mus-Qs-combined} for our four fiducial examples.

\begin{figure*}[t!]
\includegraphics[width=0.48\textwidth]{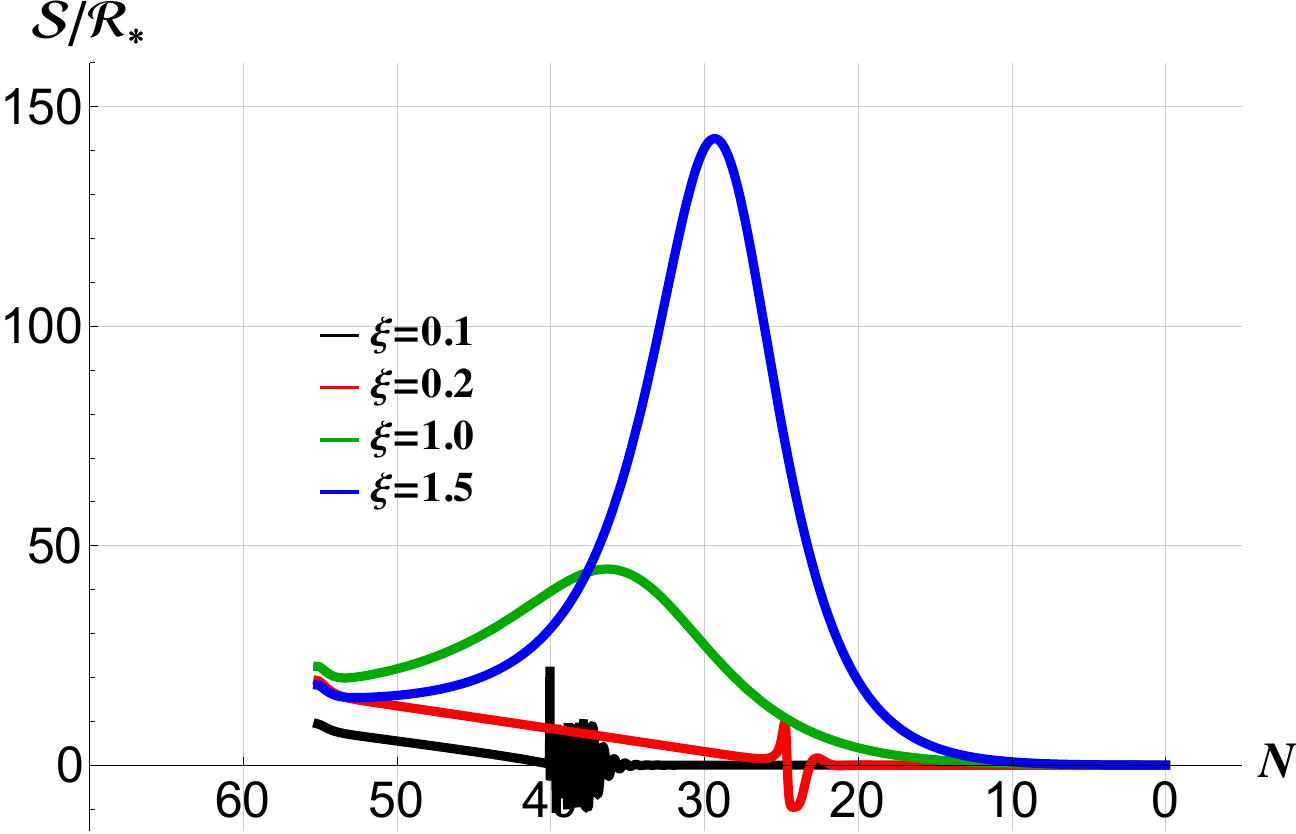}
\includegraphics[width=0.48\textwidth]{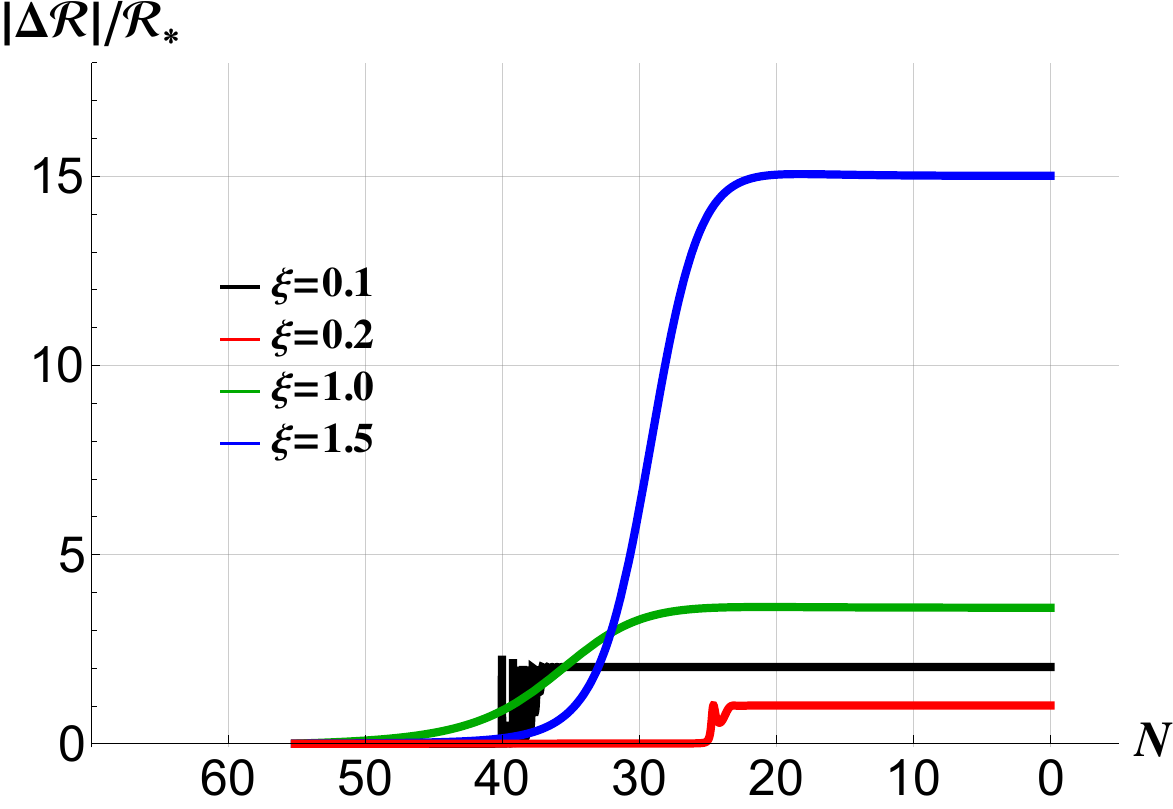}
\caption{ Evolution of the isocurvature and curvature perturbations between the time of Hubble crossing and the end of inflation, for modes with comoving $k_*$ corresponding to the CMB pivot scale. There is ${\cal O}(1)$ growth of ${\cal R} (k, t)$ in the $\xi=0.1$ (black) and $\xi=0.2$ (red) cases, whereas $|\Delta {\cal R} (k, t)| / {\cal R}_* (k) =3.58$ for $\xi = 1.0$ (green) and $|\Delta {\cal R} (k, t)| / {\cal R}_* (k) =15.01$ for $\xi = 1.5$ (blue). (Here ${\cal R}_* (k)$ is the constant amplitude of the mode following Hubble crossing, before the turn.) Initial conditions and other parameters for the curves shown here are listed in Appendix \ref{app:params}. Each curve extends to the end of inflation, when $\epsilon(N_{\rm end})=1$. }
\label{fig:SRplot}
\end{figure*}

From the behavior of modes $Q_s (k, t)$ we may understand the evolution of ${\cal S} (k, t)$ and, in turn, the effects on the curvature perturbations ${\cal R} (k, t)$. The modes $Q_s (k, t)$ and ${\cal S} (k, t)$ are related via Eq.~(\ref{eq:Sdef}), and hence the amplitude of modes ${\cal S} (k, t)$ is sensitive both to the tachyonic growth of $Q_s (k, t)$ prior to the field-space turn, as well as to the evolution of the slow-roll parameter $\epsilon (t)$, given the relation $(H / \dot{\sigma}) = 1 / (\sqrt{ 2 \epsilon (t)} \, M_{\rm pl})$. For each of the cases of interest, we find marked growth of modes ${\cal S} (k, t)$ following the turn, despite the (temporary) growth of $\epsilon (t)$ at the turn, as shown in the left panel of Fig.~\ref{fig:SRplot}. In each of these cases, the amplitude of the isocurvature modes ${\cal S} (k, t)$ rapidly decays following the turn, due to the large mass ($\mu_s^2 / H^2 \gg 1$) during the second inflation phase.

From Eq.~(\ref{eq:dotRS}), we may evaluate the effect on the curvature perturbations ${\cal R} (k, t)$. In general, the change to modes ${\cal R} (k, t)$ for $k \ll aH$ is given by
\begin{equation}
\Delta {\cal R} (k, t) = \int_{t_*}^{t} {\rm d}t'\,  2 \omega (t')\, {\cal S} (k, t') \, .    \label{eq:DeltaRdef}
\end{equation}
The growth and subsequent decay of ${\cal S}$ is imprinted on $\cal{R}$ as a rapid increase around the time when the turn rate $\omega$ is peaked. Evaluating Eq.~(\ref{eq:DeltaRdef}) numerically for the mode $k_*$ corresponding to the CMB pivot scale for the cases shown in Figs.~\ref{fig:omegaepsplot} - \ref{fig:SRplot}, we find $|\Delta {\cal R} (k_*, t_{\rm end})| \simeq 2.04 \, {\cal R}_*$ for $\xi = 0.1$, $|\Delta {\cal R} (k_*, t_{\rm end})| \simeq 1.01 \, {\cal R}_*$ for $\xi = 0.2$, $|\Delta {\cal R} (k_*, t_{\rm end})| \simeq 3.58 \, {\cal R}_*$ for $\xi = 1$, and $|\Delta {\cal R} (k_*, t_{\rm end})| \simeq 15.01 \, {\cal R}_*$ for $\xi=1.5$, where $t_{\rm end}$ is the end of inflation (at the end of the axion-inflation phase), and ${\cal R}_*$ is the amplitude ${\cal R} (k_*, t_*)$ at the time the mode crossed outside the Hubble radius, during the radial-inflation phase.  (The turn-rate $\omega (t)$ depends on several parameters, and hence the resulting amplification $\vert \Delta {\cal R} (k_*, t_{\rm end} ) \vert$ that we find for our four fiducial cases does not scale monotonically with $\xi$.)

The power spectrum of curvature perturbations ${\cal P}_{\cal R} (k, t) \equiv k^3 \vert {\cal R} (k, t) \vert^2 / (2 \pi^2)$ may be evaluated at the end of inflation:
\be
\label{eq:Pzetafinal}
{\cal P}_{\cal R} (k, N_{\rm end}) = \left[ 1 + \left(\frac{\Delta {\cal R} (k, N_{\rm end}) }{{\cal R}_* (k)}\right)^2 \right] {\cal P}_{\cal R}^{(0)} (k) ,
\ee
where ${\cal P}_{\cal R}^{(0)} (k)$ is given by
\be
{\cal P}_{\cal R}^{(0)} (k) = \frac{H_*^2}{8 \pi^2 M_{\rm pl}^2 \epsilon_*}
\ee
and is evaluated at the time that the CMB pivot scale crosses outside the Hubble radius. We numerically compute the power spectrum and evaluate the spectral index $n_s = 1 + d \log {\cal P}_{\cal R} / d \log k$ and tensor-to-scalar ratio, which will be discussed in the following section.

Similarly, we may evaluate the power spectrum of isocurvature perturbations at late times, defined as,
\begin{equation}
    {\cal P}_{\cal S} (k,N) \equiv \frac{k^3}{2 \pi^2} |{\cal S}(k, N)|^2  ,
    \label{eq:PS}
\end{equation}
evaluated at the end of inflation $N_{\rm end}$. Around $N_*$, we expect ${\cal P}_{\cal S} (k, N_*) \simeq {\cal P}_{\cal R} (k, N_*) = {\cal P}_{\cal R}^{(0)} (k)$, given the similar evolution of $Q_s (k, t)$ and $Q_\sigma (k, t)$. Following the turn, the modes $Q_s (k, t)$ decay as $a^{-3/2} (t)$ and ${\cal S} (k, t) \rightarrow 0$, as shown in Fig.~\ref{fig:SRplot}. This ensures that only a negligible amplitude of isocurvature perturbations remains at the end of inflation. We compare to observational constraints in the following section. 

Finally, we note that whereas in the examples studied here the amplification of isocurvature modes around the turn stems largely from tachyonic amplification, one may expect that for other choices of parameters, the change of ${\cal S} (k, t)$ could be further enhanced by parametric resonance as the radial field $\varphi$ rapidly oscillates around the minimum of its potential, $\varphi \simeq v$. Those rapid oscillations would contribute quasi-periodic variations to $\mu_s^2 (t)$, akin to the resonances studied in preheating in similar multifield models with nonminimal couplings. In general, the strength of such resonances grows with $\xi$ \cite{DeCross:2015uza,DeCross:2016fdz,DeCross:2016cbs,Nguyen:2019kbm,vandeVis:2020qcp}, and may become a significant factor for the growth of ${\cal S} (k, t)$ around $t_{\rm turn}$ in this model for $\xi \gg 1$. We leave this interesting possibility for future research.

\section{Predictions And Implications For Next-Generation CMB Experiments}
\label{sec:observables}

We now turn to the observable predictions of the model developed here. The hallmark observables of any inflationary model are the predictions for the scalar spectral index $n_s$ and for the tensor-to-scalar ratio $r$. The spectral index in the model proposed here depends upon the evolution of the slow-roll parameters as imprinted on the isocurvature perturbations that are later amplified and converted to (adiabatic) curvature perturbations. We compute $n_s$ numerically for all examples considered here.

A striking feature of the model is the implication of the amplication of scalar modes for the tensor-to-scalar ratio $r$.  For the multifield axion model under consideration, the amplitude of the scalar spectrum is rescaled according to Eq.~\eqref{eq:Pzetafinal}. Meanwhile, tensor modes on long length-scales are unaffected by the turn in field space. One can appreciate this from the equation of motion for tensor modes, at linear order in perturbations (see, e.g., Ref.~\cite{Baumann:2009ds}):
\begin{equation}
    u_k '' + \left( k^2 - \frac{a''}{a}\right)u_k = 0 \, ,
    \label{heom}
\end{equation}
where primes denote derivatives with respect to conformal time, $d\tau = dt / a$, and $u_k \equiv a h_k$, where $a$ is the scale factor and $h_k$ is the tensor mode function. Details of the evolution of the background system (including the turn) enter Eq.~(\ref{heom}) only through $a(t)$. In the limit of long-wavelength perturbations, $k\rightarrow 0$, Eq.~(\ref{heom}) is solved by $u_k \propto a$, and hence constant $h_k$. This indicates that $h_k$ is unaffected by any background evolution that occurs long after the tensor mode with wavenumber $k$ has exited the horizon. Given the lack of enhancement of tensor modes, the amplitude relative to the scalar perturbations, Eq.~\eqref{eq:Pzetafinal}, is suppressed:
\be
r \rightarrow \frac{r}{1 + (\Delta {\cal R}/{\cal R}_* )^2} \, ,
\label{radjust1}
\ee
where $\Delta {\cal R} / {\cal R}_*$ is shorthand for $\Delta {\cal R} (k, t_{\rm end} ) / {\cal R}_* (k)$, and we consider modes with comoving wavenumber near the CMB pivot scale $k_*$. For the parameters considered in the previous section, this suppression amounts to roughly a factor of $5.16$ for $\xi=0.1$, $2.02$ for $\xi=0.2$, $13.83$ for $\xi=1$, and $226.3$ for $\xi=1.5$.  The predicted value of $r$, used in Fig.~\ref{fig:SO} below, is taken to be the usual slow-roll approximation, $r \approx 16 \epsilon_*$, modified by the suppression factor.

Future experiments will significantly improve the sensitivity to $r$. With this in mind, in Fig.~\ref{fig:SO} we compare the predictions of our four fiducial models, for fixed $N_*=55$, to the Simons Observatory constraints, using the sensitivity forecasts in Ref.~\cite{Ade:2018sbj}. In addition, the recent measurement of gravitational waves by the NANOGrav collaboration \cite{NANOGrav:2023hvm} motivates the exploration of inflation models with exponentially small values of $r$. We will present an analysis of the running of the spectral index in upcoming work, and here focus solely on these four dots in the $(n_s,r)$ plane. 

The grey contours in Fig.~\ref{fig:SO} correspond to observations by {\it Planck} 2018 \cite{Akrami:2018odb}, while the orange and red contours correspond to expected sensitivities of the Simons Observatory.  The dots correspond to predictions from the multifield axion model of Eq.~(\ref{SE}) with $N_* = 55$. The four examples have the same spectral index $n_s=0.965$ and $A_s \equiv {\cal P}_{\cal R} (k_*,N_{\rm end}) = 2.1 \times 10^{-9}$, and differing tensor-to-scalar ratios, $r=1.8 \times 10^{-2}$ { (for $\xi = 0.1$)}, $r = 1.1\times10^{-2}$ { (for $\xi = 0.2$)}, $r = 1.1 \times 10^{-3}$ { (for $\xi = 1.0$)}, and $r = 1.1\times 10^{-4}$ { for $\xi = 1.5$).} From Fig.~\ref{fig:SO} one can appreciate that predictions from the model remain well inside the forecast contours, and in particular for $\xi=1.0$ and $\xi=1.5$ the models lie inside the forecast $1\sigma$ contour of the enhanced sensitivities for the Simons Observatory.

\begin{figure}[h!]
\includegraphics[width=0.45\textwidth]{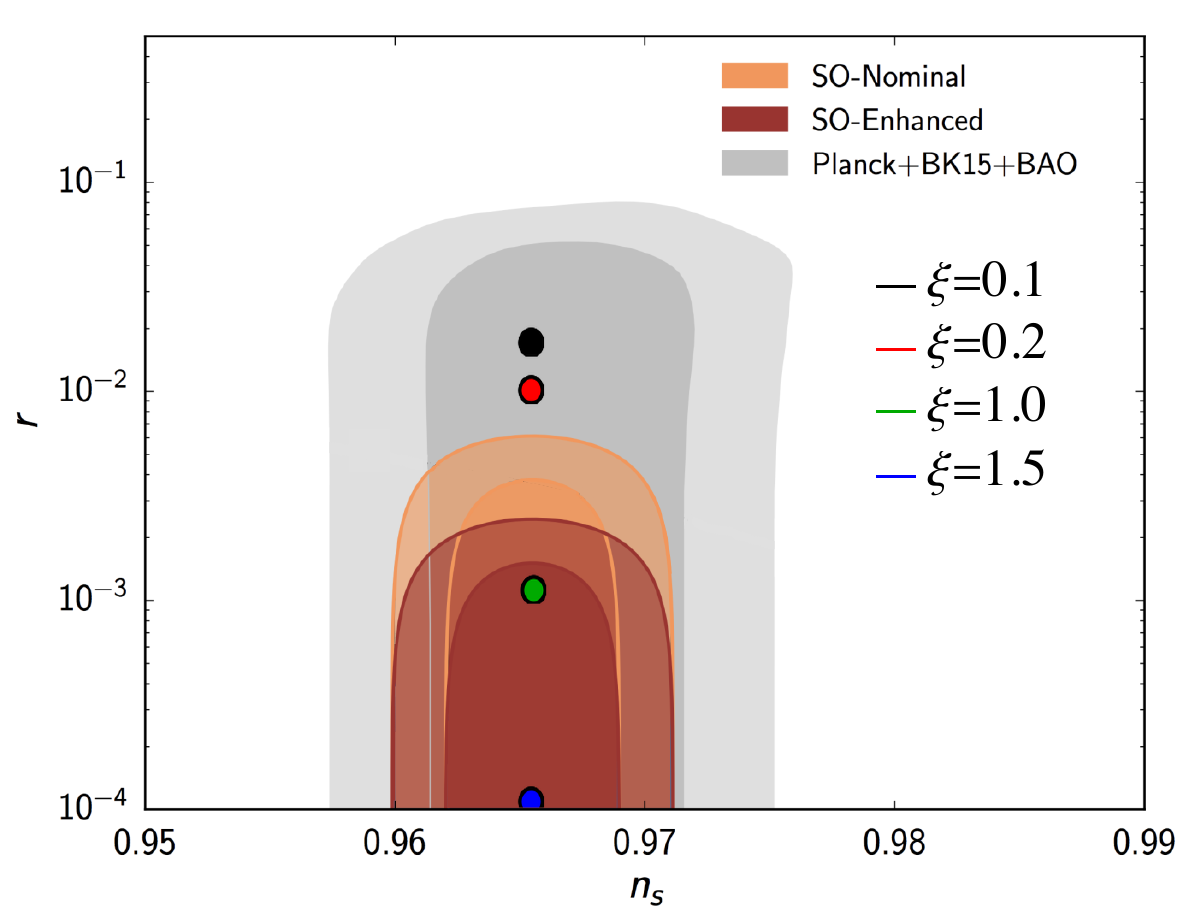}
\caption{Simons Observatory forecast for the $n_s-r$ plane, and constraints on inflation. Image adapted from the Simons Observatory forecast of Ref.~\cite{Ade:2018sbj}. Predictions for $n_s$ and $r$ with  $\xi = 0.1$ (black), $\xi = 0.2$  (red), $\xi = 1.0$  (green), and $\xi = 1.5$ (blue), for fixed $N_*=55$  and all other parameters given in Appendix \ref{app:params}. Parameters are chosen to match $n_s=0.965$ and $A_s=2.1\times10^{-9}$ in all cases. }
\label{fig:SO}
\end{figure}

As noted in the previous section, the multifield axion model under study produces isocurvature perturbations during inflation as well as (adiabatic) curvature perturbations. The CMB places tight constraints on primordial isocurvature perturbations among $\Lambda$CDM components. Whereas inflationary isocurvature perturbations are a necessary condition for observable isocurvature perturbations within the CMB, they are not a sufficient condition. An observable isocurvature fraction also requires that the inflationary isocurvature perturbations be transferred to $\Lambda$CDM components. In this context, we may calculate the maximum possible isocurvature fraction that could be observed within the CMB, given the amplification of primordial isocurvature perturbations during inflation. The observed isocurvature fraction will then be bounded by
\be
\beta_{\rm iso} (k) \leq \frac{{\cal P}_{\cal S} (k, N_{\rm end}) }{{\cal P}_{\cal R} (k, N_{\rm end}) + {\cal P}_{\cal S} (k, N_{\rm end}) } \, .
\ee
For the multifield axion model, ${\cal P}_{\cal S} (k, N_{\rm end})$ and ${\cal P}_{\cal R} (k, N_{\rm end})$ are given by Eqs.~\eqref{eq:PS} and \eqref{eq:Pzetafinal}, respectively. For { our fiducial sets of parameters} we find $\beta_{\rm iso} \simeq 10^{-50}$ { (for $\xi = 0.1$)}, $\beta_{\rm iso} \simeq 10^{-32}$ { (for $\xi = 0.2$)}, $\beta_{\rm iso} \simeq 10^{-9}$ { (for $\xi = 1.0$)}, and $\beta_{\rm iso} \simeq 10^{-8}$ { (for $\xi = 1.5$).} Such values are many orders of magnitude below the present bounds from CMB observations \cite{Akrami:2018odb}.

Finally we turn to non-Gaussianity on CMB scales, as parameterized by the amplitude of the bispectrum, $f_{\rm NL}$. As we found in Sec.~\ref{sec:inflation}, for reasonable choices of parameters and initial conditions for this model, the energy density and hence dynamics of inflation are initially dominated by the radial field. That means that the fluctuations on CMB scales crossed outside the Hubble radius within the radial-inflation phase, prior to the fast turn and onset of axion inflation. Since both curvature perturbations and isocurvature perturbations were effectively massless at the time they first crossed outside the Hubble radius, no non-Gaussianity should be generated in this model beyond the usual, slow-roll suppressed contribution. 

The lack of sizeable non-Gaussianity in ultra-light isocurvature scenarios has been emphasized in many recent works, e.g., Refs.~\cite{Achucarro:2016fby,Achucarro:2019pux,Achucarro:2019lgo,Achucarro:2019mea}. We leave a complete calculation of $f_{\rm NL}$, which although unobservably small may differ in form from the single-field result \cite{Achucarro:2019pux}, to future work.

An additional feature of the fast turn in field space that may yield an observable signal would be a short-lived oscillation of $a(t)$ associated with the rapid turn. Such ``primordial standard clock" features \cite{Chen:2014joa,Chen:2014cwa,Chen:2015lza,Chen:2016qce,Domenech:2019cyh,Braglia:2020fms} might be observable in the high multipoles of the CMB spectrum, and remain the subject of further research.

\section{Implications for Gauge Field Production}
\label{sec:gaugefields}

The canonical interaction of an axion field is with gauge fields \cite{PhysRevD.98.030001}. The interaction can be written in component form as
\begin{equation}
\label{phiFFtilde}
S_{\rm int} \, = \frac{\hat{\alpha}}{4} \int d^4 x \sqrt{-g} \, \vartheta F_{\mu \nu} \tilde{F}^{\mu \nu} \, ,
\end{equation}
where $\vartheta$ is the axion field, $\hat{\alpha}$ is the interaction strength (not to be confused with the parameter $\alpha$ defined in Eq.~(\ref{eq:alpha})), $F_{\mu \nu} \, = \, \partial_\mu A_{\nu } - \partial_\nu A_{\mu}$ is the gauge field strength, and $\tilde{F}^{\mu \nu} \equiv \epsilon^{\mu \nu \rho \sigma}F_{\rho \sigma}$. Despite the explicit factor of $\sqrt{-g}$ appearing in Eq.~\eqref{phiFFtilde}, the interaction is actually invariant under conformal rescalings of the metric. This implies that the coupling between the axion and the gauge field is {\it not} rescaled by the conformal factor, even after performing the transformation $\tilde{g}_{\mu\nu} \rightarrow g_{\mu\nu}$.

There is a vast literature on the phenomenology of an interaction of the form given in Eq.~(\ref{phiFFtilde}) during inflation. The equation of motion for gauge-field fluctuations is (see, e.g., Refs.~\cite{Anber:2009ua,Barnaby:2011qe,Adshead:2015pva,Adshead:2016iae})
\be \label{EoMA}
\frac{d^2 {A}_{k\pm}}{d \tau^2} + \left( k^2 \pm 2k \frac{\kappa}{\tau} \right) A_{k\pm} \, = \, 0 \, , 
\ee
where $\pm$ denote the two polarizations of the gauge field, $\tau$ is conformal time ($d \tau = dt / a$), $k$ is the comoving wavenumber, and $\kappa$ is given by
\begin{equation} 
\kappa \, = \, \frac{\hat{\alpha} |\dot{\vartheta}|}{ 2 H} \, ,
\label{eq:kappa}
\end{equation}
where the dot denotes a derivative with respect to cosmic time, $t$.

\begin{figure}
\centering
\includegraphics[width=0.47\textwidth]{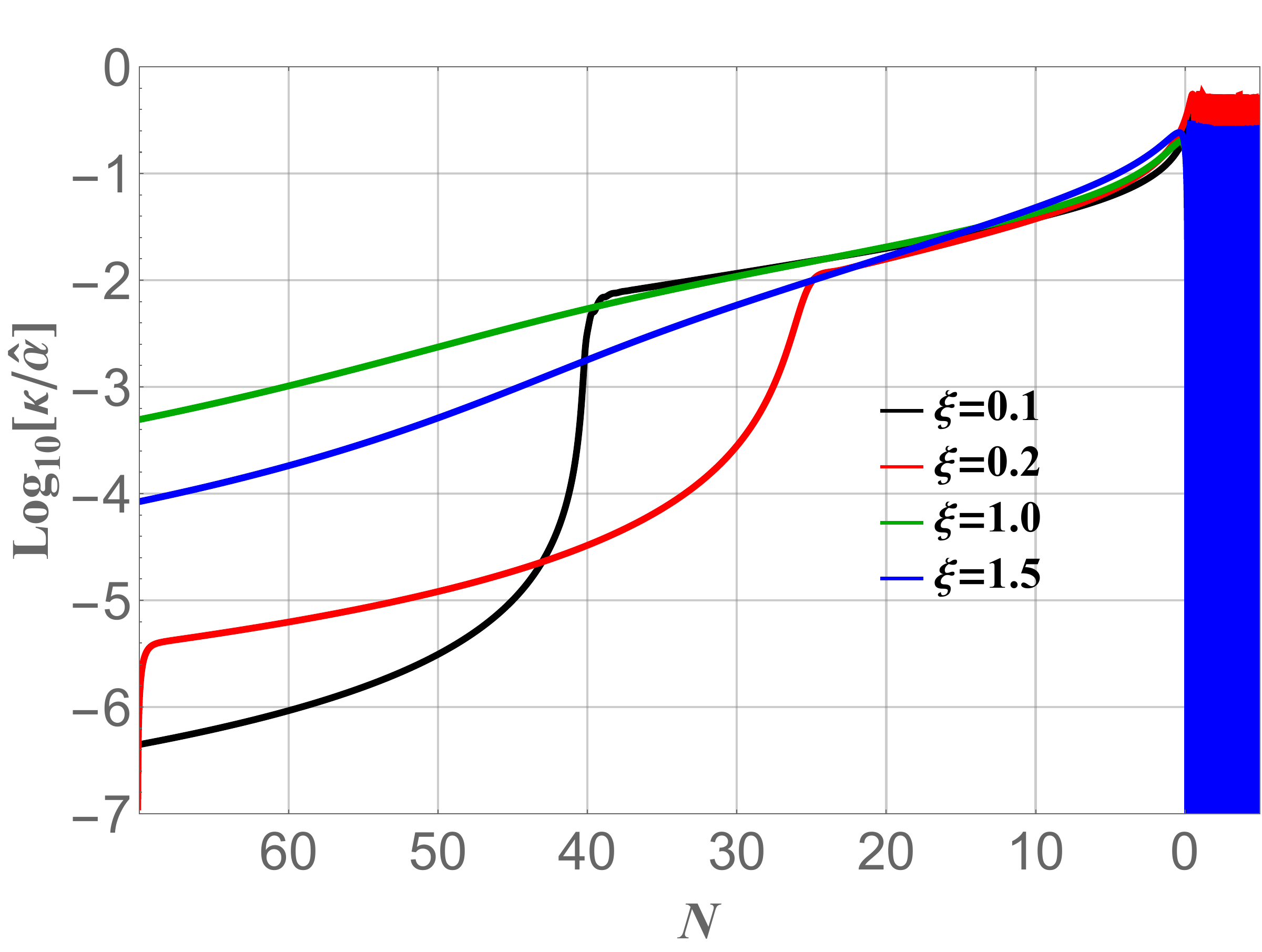}
\caption{Evolution of the interaction strength of the axion field to gauge fields, $\kappa$, normalized by the coupling constant $\hat{\alpha}$, for  $\xi=0.1$ (black), $\xi$=0.2 (red), $\xi$=1.0 (green), and $\xi=1.5$ (blue), and other parameters given in Appendix \ref{app:params}. The interaction strength during the axion-inflation phase is orders of magnitude larger than its value during the radial-inflation phase. }
\label{fig:xiplot}
\end{figure}

The equation of motion exhibits a tachyonic instability on large scales, for modes satisfying
\be
k \leq \frac{2 \kappa}{|\tau|} \, .
\label{eq:tachyonic}
\ee
Due to the $k$ dependence of the effective mass in Eq.~\eqref{EoMA}, the tachyonic instability is strongest at exactly the moment of horizon crossing, which occurs when the inequality of Eq.~(\ref{eq:tachyonic}) is saturated. This is reflected in the solution to Eq.~\eqref{EoMA}, which on large scales is given by \cite{Anber:2009ua}
\bea
\label{modefunctioninflation}
A_{k+} && \, = \,
\frac{1}{\sqrt{2k}} \left( \frac{k}{ 2 \kappa_* a H}\right)^{1/4} e^{\pi \kappa_* - 2 \kappa_* \sqrt{ 2 k /  (\kappa_* a H ) }},\\
A_{k-} && \, \simeq \, 0.
\eea
The $+$ polarization state is amplified by a factor of $e^{\pi \kappa_*}$, where $\kappa_*$ is $\kappa$ evaluated at the moment a mode $k$ entered the tachyonic regime of Eq.~\eqref{eq:tachyonic}. The amount of amplification of the mode $A_{k+}$ is therefore controlled by the parameter $\kappa$. We plot the evolution of $\kappa$, normalized by $\hat{\alpha}$, in Fig.~\ref{fig:xiplot}. For the two examples under consideration with $\xi < 1$, the interaction strength during the axion-inflation phase is orders of magnitude larger than the strength during the radial-inflation phase.

The production of gauge fields in this manner during inflation is constrained by primordial black holes \cite{Garcia-Bellido:2016dkw} and CMB non-Gaussianity. These constraints apply to modes with comoving wavenumbers $k$ that exited the horizon at relatively early times during inflation, respectively. Interestingly, neither constraint applies to modes that exited the horizon during the late stages (e.g., the last 10 e-folds) of inflation \cite{Garcia-Bellido:2016dkw}, which is precisely the regime of strong gauge-field production identified here.

In contrast, as discussed in detail in Refs.~\cite{Adshead:2019lbr,Adshead:2019igv}, there are tight constraints on the production of gauge fields after inflation, namely, during preheating. In the present model, these constraints are decoupled from the system's evolution during the phase of radial inflation. We leave a detailed study of preheating in this model to future work.

\section{Discussion}
\label{sec:discussion}

In this paper we have studied the multifield dynamics of axion inflationary models. Starting from fairly generic ingredients --- a complex scalar field, a nonminimal coupling, and the typical potential for an axion field --- we have identified an interacting two-field inflationary model. Given that axion models are inherently multifield, and that nonminimal couplings are an inevitable feature of self-interacting scalar fields in curved spacetime, this scenario captures features that we expect to be common across axion-inflation models.

The inflationary dynamics exhibit a transition between { a phase of inflation driven predominantly by the radial field and a phase wherein the axion field contributes ${\cal O}(10)\%$ to $100\%$ of the energy density of the universe.} The dynamics of the first (radial) inflation phase are controlled by the nonminimal coupling $\xi$, whereas the dynamics of the second (axion) inflation phase are set by the axion decay constant and the axion field's initial condition. The hierarchy of Hubble parameters during the two phases is determined by the hierarchy of couplings associated with the symmetry-breaking potential for the radial field and with the nonperturbative potential of the axion field. 

Cosmological perturbations in this model are readily studied using the covariant formalism of Refs.~\cite{GrootNibbelink:2000vx,GrootNibbelink:2001qt,Seery:2005gb,Langlois:2008mn,Peterson:2010np,Gong:2011uw,Kaiser:2012ak,Gong:2016qmq}, which allows for a straightforward calculation of the conversion of isocurvature perturbations into curvature perturbations, in terms of the integrated covariant turn rate of the background fields' trajectory during the inflationary evolution. From this we find an enhancement of the magnitude of the power spectrum of the primordial curvature perturbation on super-Hubble length-scales, up to ${\cal O}(10^2)$ for $\xi = 1.5$, generated by the conversion of isocurvature perturbations into curvature perturbations. We further find a very strong suppression of isocurvature perturbations at late times, easily satisfying observational bounds.

On the other hand, the tensor perturbations on long length-scales remain unaffected by the rapid turn in field space, and hence the tensor-to-scalar ratio $r$ is suppressed by the relative amplification of the scalar curvature perturbations. The predictions for observables such as the spectral index $n_s$ and the tensor-to-scalar ratio $r$ are shifted to sit well inside the forecasted sensitivity of the Simons Observatory, see Fig.~\ref{fig:SO}. This implies that values of the nonminimal coupling $\xi$ that might naively be ruled out by a non-observation of primordial gravitational waves --- such as single-field models with $\xi=1.5$, which suggest a tensor-to-scalar ratio $r \sim 3-5 \times 10^{-3}$ --- would remain consistent with observations once the multifield dynamics, including the turning trajectory in field space, are taken into account.

There are many additional avenues of interest regarding the multifield dynamics of axion inflation. First amongst these is the identification of observable signatures of the model, in lieu of the tensor-to-scalar ratio. Recent work \cite{Lorenzoni:2024krn} and numerous studies, such as Refs.~\cite{Adshead:2010mc,Hardwick:2018zry,Easther:2021rdg,Bahr-Kalus:2022prj,Martin:2024nlo,Munoz:2016owz}, suggest the {\it running} of the spectral index as a  powerful test of inflation models. It will be interesting to explore the parameter dependence of the running in the multifield axion inflation scenario.

 Our analysis has focused solely on the predictions for perturbations on CMB scales, namely long-wavelength perturbations, and we have not considered the behavior of modes that exit the horizon around the time of the fast turn. The evolution of such modes is known to generate a spike in the power spectrum, which can be relevant to the production of primordial black holes \cite{Palma:2020ejf,Fumagalli:2020adf,Aldabergenov:2020bpt}, and other phenomena, such as CMB spectral distortions \cite{Chluba:2019kpb,Josan:2009qn}. In addition, we have identified two distinct phases for the effective interaction strength between the axion field and gauge fields, which remains suppressed during the first (radial) phase of inflation but grows rapidly around the time of the turn to axion inflation. This behavior effectively decouples observable constraints on gauge-field production during inflation from constraints during the preheating era. A more detailed study of the phenomenology of gauge-field production in this model, including the generalization to non-Abelian gauge fields (for recent work see Ref.~\cite{Iarygina:2023mtj}), predictions for the spectrum of primordial black holes, the gravitational wave signature, as well as the dynamics of preheating \cite{Adshead:2015pva,McDonough:2016xvu,Lozanov:2019jff}, remain subjects for further research.

\acknowledgements

The authors thank Stephon Alexander, Mustafa Amin, Tom Giblin, Mark Hertzberg, Wayne Hu, Keisuke Inomata, Dario Lorenzoni, and Dong-Gang Wang for helpful discussions. E.M.'s work at MIT was supported in part by a Banting Fellowship from the government of Canada. This research was conducted in MIT's Center for Theoretical Physics and supported in part by the U.S.~Department of Energy under Contract No.~DE-SC0012567. E.M. is supported in part by a Discovery Grant from the Natural Sciences and Engineering Research Council of Canada, and by a New Investigator Operating Grant from Research Manitoba.\\

\appendix

\section{Field Space Quantities}
\label{app:fieldspace}

In the Einstein frame, the nonvanishing components of the field-space metric are given in Eq.~(\ref{GEIJ}). Upon using Eq.~(\ref{fdef}) for the nonminimal coupling function $f (\varphi)$, we find the nonvanishing components of the Christoffel symbols associated with ${\cal G}_{IJ}^{(E)}$:
\begin{equation}
\begin{split}
\Gamma^{\varphi}_{\varphi \varphi}  &= \frac{\xi \varphi}{2 f C}\left[6 \xi M^2- C  \right] \, , \\ 
\Gamma^{\varphi} _{\vartheta\vartheta}  &=  - \frac{M^2 \varphi}{C} \, , \\
 \Gamma^{\vartheta} _{\vartheta \varphi}  &=\frac{ M^2}{2 f \varphi} \; \, ,
 \end{split}
\end{equation}
where we have defined the quantity
\begin{equation}
    C \equiv 2 f + 6 \xi^2 \varphi^2 \, .
\end{equation}
The Ricci scalar of the field-space manifold in the Einstein frame is given by
\be
\label{eq:RE}
\mathcal{R}_{E}(\varphi)=\frac{ F (\xi, r_\varphi) }{M_{\rm pl}^2 }
\ee
where
\be
F (\xi, r_\varphi)=\frac{4 \xi [1 + 3 \xi + (1+6\xi) r_\varphi^2]} {\ [ 1 + (1 + 6 \xi) r_\varphi^2]^2},
\ee
and
\be
r_\varphi \equiv \frac{\sqrt{\xi} \varphi}{M}.
\ee

\widetext
\section{Parameters for Examples Shown in the Figures}
\label{app:params}
\endwidetext

Fig.~\ref{fig:xi-plots}, left panel:\newline \newline $\xi=1/6, 1/8, 1/10, 1/12, 1/14$\\ $v/ M_{\rm pl}=2$\\ $10^{11} \lambda=11.94, 6.75, 4.35, 3.042, 2.25$\\ $\Lambda = 1.529 \times 10^{-3} \, M_{\rm pl}$\\ $\varphi_i/ M_{\rm pl} = 14.5, 15.5, 16.2, 16.7, 17.1$ \\ $\vartheta_i/\pi = 0.9$\\

Fig.~\ref{fig:xi-plots}, right panel:\newline \newline $\xi=100, 10, 5, 1, 1/2$\\ $v/M_{\rm pl} =\frac{9}{10 \sqrt{ \xi}}$\\ $\lambda = 4 \times 10^{-9}\, \xi^2$\\ $\Lambda = 1.934 \times 10^{-3}\,M_{\rm pl}$\\ $\varphi_i/ M_{\rm pl} = 0.90, 2.84, 3.97, 8.37, 11.1$ \\ $\vartheta_i/\pi = 1-10^{-20}, 1-5 \times 10^{-7}, 1-2 \times 10^{-5}, 0.98, 0.92$\\

Fig.~\ref{fig:Fig5plus}:\newline\newline $\xi=10,5,1,1/2$\\ $v/M_{\rm pl} = 0.3$\\ $\lambda = 4 \times 10^{-5} \xi^2$\\ $\Lambda = 0.01368$\\ $\varphi_i/M_{\rm pl} = 3, 4, 8, 1$\\ $\vartheta_i/\pi = 1-10^{-25}, 1-3\times 10^{-25}, 1-10^{-24}, 1-5 \times 10^{-21}$\\

Fig.~\ref{fig:Na-rH-plots}, left panel:\newline \newline $\xi=1$\\ $v/M_{\rm pl} =\sqrt{\frac{99}{100}}$\\ $\lambda = 4 \times 10^{-9} \xi^2$\\ $\Lambda = 2.3 \times 10^{-3} M_{\rm pl}$\\ $\varphi_i/ M_{\rm pl} = 8.5$ \\ $\vartheta_i/\pi = 0.99, 0.975, 0.95, 0.9, 0.85$\\

\vbox{%
Fig.~\ref{fig:Na-rH-plots}, right panel:\newline \newline $\xi=1$\\ $v/M_{\rm pl} =\sqrt{\frac{99}{100}}$\\ $\lambda = 4 \times 10^{-9} \xi^2$\\ $10^3 \Lambda/M_{\rm pl} =2.24, 2.17, 2.09, 2.00, 1.90$\\ $\varphi_i/ M_{\rm pl} = 8.5$ \\ $\vartheta_i/\pi = 0.95$\\}

\newpage

Examples for Figs.~\ref{fig:H-plot-perts}--\ref{fig:xiplot}:\newline

Black: \newline $\xi=0.1$\\ $v/M_{\rm pl}=3.13065$\\ $\lambda = 2.77065 \times 10^{-11}$\\ $\Lambda = 1.59876 \times 10^{-3} M_{\rm pl}$\\ $\varphi_i/ M_{\rm pl} = 15.0$ \\ $\vartheta_i/\pi = 0.91$\\

Red: \newline $\xi=0.2$\\ $v/M_{\rm pl} =2.21371$\\ $\lambda = 5.76798 \times 10^{-11}$\\ $\Lambda = 2.79268 \times 10^{-3}$\\ $\varphi_i/ M_{\rm pl} = 13.1$ \\ $\vartheta_i/\pi = 0.917$\\

Green: \newline $\xi=1.0$\\ $v/M_{\rm pl}=0.995$\\ $\lambda = 1.56945 \times 10^{-10}$\\ $\Lambda = 3.57368 \times 10^{-3}$\\ $\varphi_i/ M_{\rm pl} = 8.2$ \\ $\vartheta_i/\pi = 0.952$\\

Blue: \newline $\xi=1.5$\\ $v/M_{\rm pl}=0.808332$\\ $\lambda = 3.34294 \times 10^{-11}$\\ $\Lambda = 1.85638 \times 10^{-3}$\\ $\varphi_i/ M_{\rm pl} = 6.0$ \\ $\vartheta_i/\pi = 0.99$\\

\vspace{2cm}

\newpage

\bibliographystyle{JHEP}
\bibliography{axion-refs}

\providecommand{\href}[2]{#2}\begingroup\raggedright\begin{thebibliography}{100}

\bibitem{Peccei:1977hh}
R.~D. Peccei and H.~R. Quinn, \emph{{CP Conservation in the Presence of
  Instantons}}, \href{https://doi.org/10.1103/PhysRevLett.38.1440}{\emph{Phys.
  Rev. Lett.} {\bfseries 38} (1977) 1440}.

\bibitem{Wilczek:1977pj}
F.~Wilczek, \emph{{Problem of Strong $P$ and $T$ Invariance in the Presence of
  Instantons}}, \href{https://doi.org/10.1103/PhysRevLett.40.279}{\emph{Phys.
  Rev. Lett.} {\bfseries 40} (1978) 279}.

\bibitem{Weinberg:1977ma}
S.~Weinberg, \emph{{A New Light Boson?}},
  \href{https://doi.org/10.1103/PhysRevLett.40.223}{\emph{Phys. Rev. Lett.}
  {\bfseries 40} (1978) 223}.

\bibitem{Freese:1990rb}
K.~Freese, J.~A. Frieman and A.~V. Olinto, \emph{{Natural inflation with pseudo
  - Nambu-Goldstone bosons}},
  \href{https://doi.org/10.1103/PhysRevLett.65.3233}{\emph{Phys. Rev. Lett.}
  {\bfseries 65} (1990) 3233}.

\bibitem{McAllister:2008hb}
L.~McAllister, E.~Silverstein and A.~Westphal, \emph{{Gravity Waves and Linear
  Inflation from Axion Monodromy}},
  \href{https://doi.org/10.1103/PhysRevD.82.046003}{\emph{Phys. Rev.}
  {\bfseries D82} (2010) 046003}
  [\href{https://arxiv.org/abs/0808.0706}{{\ttfamily 0808.0706}}].

\bibitem{Silverstein:2008sg}
E.~Silverstein and A.~Westphal, \emph{{Monodromy in the CMB: Gravity Waves and
  String Inflation}},
  \href{https://doi.org/10.1103/PhysRevD.78.106003}{\emph{Phys. Rev.}
  {\bfseries D78} (2008) 106003}
  [\href{https://arxiv.org/abs/0803.3085}{{\ttfamily 0803.3085}}].

\bibitem{Preskill:1982cy}
J.~Preskill, M.~B. Wise and F.~Wilczek, \emph{{Cosmology of the Invisible
  Axion}}, \href{https://doi.org/10.1016/0370-2693(83)90637-8}{\emph{Phys.
  Lett.} {\bfseries B120} (1983) 127}.

\bibitem{Abbott:1982af}
L.~F. Abbott and P.~Sikivie, \emph{{A Cosmological Bound on the Invisible
  Axion}}, \href{https://doi.org/10.1016/0370-2693(83)90638-X}{\emph{Phys.
  Lett.} {\bfseries B120} (1983) 133}.

\bibitem{Dine:1982ah}
M.~Dine and W.~Fischler, \emph{{The Not So Harmless Axion}},
  \href{https://doi.org/10.1016/0370-2693(83)90639-1}{\emph{Phys. Lett.}
  {\bfseries B120} (1983) 137}.

\bibitem{Barnaby:2011qe}
N.~Barnaby, E.~Pajer and M.~Peloso, \emph{{Gauge Field Production in Axion
  Inflation: Consequences for Monodromy, non-Gaussianity in the CMB, and
  Gravitational Waves at Interferometers}},
  \href{https://doi.org/10.1103/PhysRevD.85.023525}{\emph{Phys.\ Rev.\ D}
  {\bfseries 85} (2012) 023525}
  [\href{https://arxiv.org/abs/1110.3327}{{\ttfamily 1110.3327}}].

\bibitem{Svrcek:2006yi}
P.~Svrcek and E.~Witten, \emph{{Axions In String Theory}},
  \href{https://doi.org/10.1088/1126-6708/2006/06/051}{\emph{JHEP} {\bfseries
  06} (2006) 051} [\href{https://arxiv.org/abs/hep-th/0605206}{{\ttfamily
  hep-th/0605206}}].

\bibitem{Arvanitaki:2009fg}
A.~Arvanitaki, S.~Dimopoulos, S.~Dubovsky, N.~Kaloper and J.~March-Russell,
  \emph{{String Axiverse}},
  \href{https://doi.org/10.1103/PhysRevD.81.123530}{\emph{Phys. Rev.}
  {\bfseries D81} (2010) 123530}
  [\href{https://arxiv.org/abs/0905.4720}{{\ttfamily 0905.4720}}].

\bibitem{Cicoli:2012sz}
M.~Cicoli, M.~Goodsell and A.~Ringwald, \emph{{The type IIB string axiverse and
  its low-energy phenomenology}},
  \href{https://doi.org/10.1007/JHEP10(2012)146}{\emph{JHEP} {\bfseries 10}
  (2012) 146} [\href{https://arxiv.org/abs/1206.0819}{{\ttfamily 1206.0819}}].

\bibitem{Maleknejad:2022gyf}
A.~Maleknejad and E.~McDonough, \emph{{Ultralight pion and superheavy baryon
  dark matter}}, \href{https://doi.org/10.1103/PhysRevD.106.095011}{\emph{Phys.
  Rev. D} {\bfseries 106} (2022) 095011}
  [\href{https://arxiv.org/abs/2205.12983}{{\ttfamily 2205.12983}}].

\bibitem{Alexander:2023wgk}
S.~Alexander, H.~Gilmer, T.~Manton and E.~McDonough,
  \emph{{\ensuremath{\pi}-axion and \ensuremath{\pi}-axiverse of dark QCD}},
  \href{https://doi.org/10.1103/PhysRevD.108.123014}{\emph{Phys. Rev. D}
  {\bfseries 108} (2023) 123014}
  [\href{https://arxiv.org/abs/2304.11176}{{\ttfamily 2304.11176}}].

\bibitem{Alexander:2024nvi}
S.~Alexander, T.~Manton and E.~McDonough, \emph{{The Field Theory Axiverse}},
  \href{https://arxiv.org/abs/2404.11642}{{\ttfamily 2404.11642}}.

\bibitem{Wands:2007bd}
D.~Wands, \emph{{Multiple field inflation}},
  \href{https://doi.org/10.1007/978-3-540-74353-8\_8}{\emph{Lect. Notes Phys.}
  {\bfseries 738} (2008) 275}
  [\href{https://arxiv.org/abs/astro-ph/0702187}{{\ttfamily
  astro-ph/0702187}}].

\bibitem{Chen:2010xka}
X.~Chen, \emph{{Primordial Non-Gaussianities from Inflation Models}},
  \href{https://doi.org/10.1155/2010/638979}{\emph{Adv. Astron.} {\bfseries
  2010} (2010) 638979} [\href{https://arxiv.org/abs/1002.1416}{{\ttfamily
  1002.1416}}].

\bibitem{Gong:2016qmq}
J.-O. Gong, \emph{{Multi-field inflation and cosmological perturbations}},
  \href{https://doi.org/10.1142/S021827181740003X}{\emph{Int. J. Mod. Phys. D}
  {\bfseries 26} (2016) 1740003}
  [\href{https://arxiv.org/abs/1606.06971}{{\ttfamily 1606.06971}}].

\bibitem{Chernikov:1968zm}
N.~Chernikov and E.~Tagirov, \emph{{Quantum theory of scalar fields in de
  Sitter space-time}}, {\emph{Ann. Inst. H. Poincare Phys. Theor. A} {\bfseries
  9} (1968) 109}.

\bibitem{Callan:1970ze}
J.~{Callan}, Curtis~G., S.~R. {Coleman} and R.~{Jackiw}, \emph{{A New improved
  energy - momentum tensor}},
  \href{https://doi.org/10.1016/0003-4916(70)90394-5}{\emph{Annals Phys.}
  {\bfseries 59} (1970) 42}.

\bibitem{Bunch:1980br}
T.~Bunch, P.~Panangaden and L.~Parker, \emph{{On renormalization of $\lambda
  \phi^4$ field theory in curved space-time, I}},
  \href{https://doi.org/10.1088/0305-4470/13/3/022}{\emph{J. Phys. A}
  {\bfseries 13} (1980) 901}.

\bibitem{Bunch:1980bs}
T.~Bunch and P.~Panangaden, \emph{{On renormalization of $\lambda \phi^4$ field
  theory in curved space-time, II}},
  \href{https://doi.org/10.1088/0305-4470/13/3/023}{\emph{J. Phys. A}
  {\bfseries 13} (1980) 919}.

\bibitem{Birrell:1982ix}
N.~Birrell and P.~Davies, \emph{{Quantum Fields in Curved Space}}. Cambridge
  Univ. Press, Cambridge, UK, 1982,
  \href{https://doi.org/10.1017/CBO9780511622632}{10.1017/CBO9780511622632}.

\bibitem{Odintsov:1990mt}
S.~D. Odintsov, \emph{{Renormalization Group, Effective Action and Grand
  Unification Theories in Curved Space-time}}, {\emph{Fortsch. Phys.}
  {\bfseries 39} (1991) 621}.

\bibitem{Buchbinder:1992rb}
I.~Buchbinder, S.~Odintsov and I.~Shapiro, \emph{{Effective action in quantum
  gravity}}. Taylor and Francis, New York, 1992.

\bibitem{Parker:2009uva}
L.~E. Parker and D.~Toms, \emph{{Quantum Field Theory in Curved Spacetime}:
  {Quantized Field and Gravity}}. Cambridge University Press, New York, 2009,
  \href{https://doi.org/10.1017/CBO9780511813924}{10.1017/CBO9780511813924}.

\bibitem{Markkanen:2013nwa}
T.~Markkanen and A.~Tranberg, \emph{{A Simple Method for One-Loop
  Renormalization in Curved Space-Time}},
  \href{https://doi.org/10.1088/1475-7516/2013/08/045}{\emph{JCAP} {\bfseries
  08} (2013) 045} [\href{https://arxiv.org/abs/1303.0180}{{\ttfamily
  1303.0180}}].

\bibitem{Linde:1991km}
A.~D. Linde, \emph{{Axions in inflationary cosmology}},
  \href{https://doi.org/10.1016/0370-2693(91)90130-I}{\emph{Phys. Lett. B}
  {\bfseries 259} (1991) 38}.

\bibitem{Linde:1993cn}
A.~D. Linde, \emph{{Hybrid inflation}},
  \href{https://doi.org/10.1103/PhysRevD.49.748}{\emph{Phys. Rev. D} {\bfseries
  49} (1994) 748} [\href{https://arxiv.org/abs/astro-ph/9307002}{{\ttfamily
  astro-ph/9307002}}].

\bibitem{Copeland:1994vg}
E.~J. Copeland, A.~R. Liddle, D.~H. Lyth, E.~D. Stewart and D.~Wands,
  \emph{{False vacuum inflation with Einstein gravity}},
  \href{https://doi.org/10.1103/PhysRevD.49.6410}{\emph{Phys. Rev. D}
  {\bfseries 49} (1994) 6410}
  [\href{https://arxiv.org/abs/astro-ph/9401011}{{\ttfamily
  astro-ph/9401011}}].

\bibitem{Linde:2018hmx}
A.~Linde, D.-G. Wang, Y.~Welling, Y.~Yamada and A.~Ach\'{u}carro,
  \emph{{Hypernatural inflation}},
  \href{https://doi.org/10.1088/1475-7516/2018/07/035}{\emph{JCAP} {\bfseries
  07} (2018) 035} [\href{https://arxiv.org/abs/1803.09911}{{\ttfamily
  1803.09911}}].

\bibitem{Aldabergenov:2020bpt}
Y.~Aldabergenov, A.~Addazi and S.~V. Ketov, \emph{{Primordial black holes from
  modified supergravity}},  \href{https://arxiv.org/abs/2006.16641}{{\ttfamily
  2006.16641}}.

\bibitem{Christodoulidis:2018qdw}
P.~Christodoulidis, D.~Roest and E.~I. Sfakianakis, \emph{{Angular inflation in
  multi-field $\alpha$-attractors}},
  \href{https://doi.org/10.1088/1475-7516/2019/11/002}{\emph{JCAP} {\bfseries
  11} (2019) 002} [\href{https://arxiv.org/abs/1803.09841}{{\ttfamily
  1803.09841}}].

\bibitem{GrootNibbelink:2000vx}
S.~Groot~Nibbelink and B.~van Tent, \emph{{Density perturbations arising from
  multiple field slow roll inflation}},
  \href{https://arxiv.org/abs/hep-ph/0011325}{{\ttfamily hep-ph/0011325}}.

\bibitem{GrootNibbelink:2001qt}
S.~Groot~Nibbelink and B.~van Tent, \emph{{Scalar perturbations during multiple
  field slow-roll inflation}},
  \href{https://doi.org/10.1088/0264-9381/19/4/302}{\emph{Class. Quant. Grav.}
  {\bfseries 19} (2002) 613}
  [\href{https://arxiv.org/abs/hep-ph/0107272}{{\ttfamily hep-ph/0107272}}].

\bibitem{Seery:2005gb}
D.~Seery and J.~E. Lidsey, \emph{{Primordial non-Gaussianities from
  multiple-field inflation}},
  \href{https://doi.org/10.1088/1475-7516/2005/09/011}{\emph{JCAP} {\bfseries
  09} (2005) 011} [\href{https://arxiv.org/abs/astro-ph/0506056}{{\ttfamily
  astro-ph/0506056}}].

\bibitem{Langlois:2008mn}
D.~Langlois and S.~Renaux-Petel, \emph{{Perturbations in generalized
  multi-field inflation}},
  \href{https://doi.org/10.1088/1475-7516/2008/04/017}{\emph{JCAP} {\bfseries
  04} (2008) 017} [\href{https://arxiv.org/abs/0801.1085}{{\ttfamily
  0801.1085}}].

\bibitem{Peterson:2010np}
C.~M. Peterson and M.~Tegmark, \emph{{Testing Two-Field Inflation}},
  \href{https://doi.org/10.1103/PhysRevD.83.023522}{\emph{Phys. Rev. D}
  {\bfseries 83} (2011) 023522}
  [\href{https://arxiv.org/abs/1005.4056}{{\ttfamily 1005.4056}}].

\bibitem{Gong:2011uw}
J.-O. Gong and T.~Tanaka, \emph{{A covariant approach to general field space
  metric in multi-field inflation}},
  \href{https://doi.org/10.1088/1475-7516/2012/02/E01}{\emph{JCAP} {\bfseries
  03} (2011) 015} [\href{https://arxiv.org/abs/1101.4809}{{\ttfamily
  1101.4809}}].

\bibitem{Kaiser:2012ak}
D.~I. Kaiser, E.~A. Mazenc and E.~I. Sfakianakis, \emph{{Primordial Bispectrum
  from Multifield Inflation with Nonminimal Couplings}},
  \href{https://doi.org/10.1103/PhysRevD.87.064004}{\emph{Phys.\ Rev.\ D}
  {\bfseries 87} (2013) 064004}
  [\href{https://arxiv.org/abs/1210.7487}{{\ttfamily 1210.7487}}].

\bibitem{Renaux-Petel:2015mga}
S.~Renaux-Petel and K.~Turzy\'nski, \emph{{Geometrical Destabilization of
  Inflation}},
  \href{https://doi.org/10.1103/PhysRevLett.117.141301}{\emph{Phys. Rev. Lett.}
  {\bfseries 117} (2016) 141301}
  [\href{https://arxiv.org/abs/1510.01281}{{\ttfamily 1510.01281}}].

\bibitem{Christodoulidis:2019mkj}
P.~Christodoulidis, D.~Roest and E.~I. Sfakianakis, \emph{{Attractors,
  Bifurcations and Curvature in Multi-field Inflation}},
  \href{https://arxiv.org/abs/1903.03513}{{\ttfamily 1903.03513}}.

\bibitem{Christodoulidis:2019jsx}
P.~Christodoulidis, D.~Roest and E.~I. Sfakianakis, \emph{{Scaling attractors
  in multi-field inflation}},
  \href{https://doi.org/10.1088/1475-7516/2019/12/059}{\emph{JCAP} {\bfseries
  12} (2019) 059} [\href{https://arxiv.org/abs/1903.06116}{{\ttfamily
  1903.06116}}].

\bibitem{Fumagalli:2020adf}
J.~Fumagalli, S.~Renaux-Petel, J.~W. Ronayne and L.~T. Witkowski,
  \emph{{Turning in the landscape: a new mechanism for generating Primordial
  Black Holes}},  \href{https://arxiv.org/abs/2004.08369}{{\ttfamily
  2004.08369}}.

\bibitem{Grocholski:2019mot}
O.~Grocholski, M.~Kalinowski, M.~Kolanowski, S.~Renaux-Petel, K.~Turzy\'nski
  and V.~Vennin, \emph{{On backreaction effects in geometrical destabilisation
  of inflation}},
  \href{https://doi.org/10.1088/1475-7516/2019/05/008}{\emph{JCAP} {\bfseries
  05} (2019) 008} [\href{https://arxiv.org/abs/1901.10468}{{\ttfamily
  1901.10468}}].

\bibitem{Garcia-Saenz:2018ifx}
S.~Garcia-Saenz, S.~Renaux-Petel and J.~Ronayne, \emph{{Primordial fluctuations
  and non-Gaussianities in sidetracked inflation}},
  \href{https://doi.org/10.1088/1475-7516/2018/07/057}{\emph{JCAP} {\bfseries
  07} (2018) 057} [\href{https://arxiv.org/abs/1804.11279}{{\ttfamily
  1804.11279}}].

\bibitem{Garcia-Saenz:2018vqf}
S.~Garcia-Saenz and S.~Renaux-Petel, \emph{{Flattened non-Gaussianities from
  the effective field theory of inflation with imaginary speed of sound}},
  \href{https://doi.org/10.1088/1475-7516/2018/11/005}{\emph{JCAP} {\bfseries
  11} (2018) 005} [\href{https://arxiv.org/abs/1805.12563}{{\ttfamily
  1805.12563}}].

\bibitem{Fumagalli:2019noh}
J.~Fumagalli, S.~Garcia-Saenz, L.~Pinol, S.~Renaux-Petel and J.~Ronayne,
  \emph{{Hyper-Non-Gaussianities in Inflation with Strongly Nongeodesic
  Motion}}, \href{https://doi.org/10.1103/PhysRevLett.123.201302}{\emph{Phys.
  Rev. Lett.} {\bfseries 123} (2019) 201302}
  [\href{https://arxiv.org/abs/1902.03221}{{\ttfamily 1902.03221}}].

\bibitem{Garcia-Saenz:2019njm}
S.~Garcia-Saenz, L.~Pinol and S.~Renaux-Petel, \emph{{Revisiting
  non-Gaussianity in multifield inflation with curved field space}},
  \href{https://doi.org/10.1007/JHEP01(2020)073}{\emph{JHEP} {\bfseries 01}
  (2020) 073} [\href{https://arxiv.org/abs/1907.10403}{{\ttfamily
  1907.10403}}].

\bibitem{Pinol:2018euk}
L.~Pinol, S.~Renaux-Petel and Y.~Tada, \emph{{Inflationary stochastic
  anomalies}}, \href{https://doi.org/10.1088/1361-6382/ab097f}{\emph{Class.
  Quant. Grav.} {\bfseries 36} (2019) 07LT01}
  [\href{https://arxiv.org/abs/1806.10126}{{\ttfamily 1806.10126}}].

\bibitem{Pinol:2020cdp}
L.~Pinol, S.~Renaux-Petel and Y.~Tada, \emph{{A manifestly covariant theory of
  multifield stochastic inflation in phase space}},
  \href{https://arxiv.org/abs/2008.07497}{{\ttfamily 2008.07497}}.

\bibitem{Fumagalli:2019ohr}
J.~Fumagalli, S.~Renaux-Petel and J.~W. Ronayne, \emph{{Higgs vacuum
  (in)stability during inflation: the dangerous relevance of de Sitter
  departure and Planck-suppressed operators}},
  \href{https://doi.org/10.1007/JHEP02(2020)142}{\emph{JHEP} {\bfseries 02}
  (2020) 142} [\href{https://arxiv.org/abs/1910.13430}{{\ttfamily
  1910.13430}}].

\bibitem{Ashoorioon:2019kcy}
A.~Ashoorioon and K.~Rezazadeh, \emph{{Non-Minimal M-flation}},
  \href{https://doi.org/10.1007/JHEP07(2020)244}{\emph{JHEP} {\bfseries 07}
  (2020) 244} [\href{https://arxiv.org/abs/1909.09806}{{\ttfamily
  1909.09806}}].

\bibitem{Palma:2020ejf}
G.~A. Palma, S.~Sypsas and C.~Zenteno, \emph{{Seeding primordial black holes in
  multi-field inflation}},  \href{https://arxiv.org/abs/2004.06106}{{\ttfamily
  2004.06106}}.

\bibitem{Achucarro:2015caa}
A.~Ach\'ucarro, V.~Atal, M.~Kawasaki and F.~Takahashi, \emph{{The two-field
  regime of natural inflation}},
  \href{https://doi.org/10.1088/1475-7516/2015/12/044}{\emph{JCAP} {\bfseries
  12} (2015) 044} [\href{https://arxiv.org/abs/1510.08775}{{\ttfamily
  1510.08775}}].

\bibitem{Alam:2024krt}
K.~Alam, K.~Dutta and N.~Jaman, \emph{{CMB Constraints on Natural Inflation
  with Gauge Field Production}},
  \href{https://arxiv.org/abs/2405.10155}{{\ttfamily 2405.10155}}.

\bibitem{Montefalcone:2022jfw}
G.~Montefalcone, V.~Aragam, L.~Visinelli and K.~Freese, \emph{{Observational
  constraints on warm natural inflation}},
  \href{https://doi.org/10.1088/1475-7516/2023/03/002}{\emph{JCAP} {\bfseries
  03} (2023) 002} [\href{https://arxiv.org/abs/2212.04482}{{\ttfamily
  2212.04482}}].

\bibitem{Salvio:2021lka}
A.~Salvio, \emph{{Natural-scalaron inflation}},
  \href{https://doi.org/10.1088/1475-7516/2021/10/011}{\emph{JCAP} {\bfseries
  10} (2021) 011} [\href{https://arxiv.org/abs/2107.03389}{{\ttfamily
  2107.03389}}].

\bibitem{Salvio:2023cry}
A.~Salvio and S.~Sciusco, \emph{{(Multi-field) natural inflation and
  gravitational waves}},
  \href{https://doi.org/10.1088/1475-7516/2024/03/018}{\emph{JCAP} {\bfseries
  03} (2024) 018} [\href{https://arxiv.org/abs/2311.00741}{{\ttfamily
  2311.00741}}].

\bibitem{Racioppi:2024zva}
A.~Racioppi and A.~Salvio, \emph{{Natural Metric-Affine Inflation}},
  \href{https://arxiv.org/abs/2403.18004}{{\ttfamily 2403.18004}}.

\bibitem{Salvio:2019wcp}
A.~Salvio, \emph{{Quasi-Conformal Models and the Early Universe}},
  \href{https://doi.org/10.1140/epjc/s10052-019-7267-5}{\emph{Eur. Phys. J. C}
  {\bfseries 79} (2019) 750}
  [\href{https://arxiv.org/abs/1907.00983}{{\ttfamily 1907.00983}}].

\bibitem{Salvio:2022mld}
A.~Salvio, \emph{{BICEP/Keck data and quadratic gravity}},
  \href{https://doi.org/10.1088/1475-7516/2022/09/027}{\emph{JCAP} {\bfseries
  09} (2022) 027} [\href{https://arxiv.org/abs/2202.00684}{{\ttfamily
  2202.00684}}].

\bibitem{RoyChoudhury:2022rva}
S.~Roy~Choudhury, S.~Hannestad and T.~Tram, \emph{{Massive neutrino
  self-interactions and inflation}},
  \href{https://doi.org/10.1088/1475-7516/2022/10/018}{\emph{JCAP} {\bfseries
  10} (2022) 018} [\href{https://arxiv.org/abs/2207.07142}{{\ttfamily
  2207.07142}}].

\bibitem{Bostan:2023ped}
N.~Bostan and S.~Roy~Choudhury, \emph{{First constraints on Non-minimally
  coupled Natural and Coleman-Weinberg inflation in the light of massive
  neutrino self-interactions and Planck+BICEP/Keck}},
  \href{https://arxiv.org/abs/2310.01491}{{\ttfamily 2310.01491}}.

\bibitem{Adams:1992bn}
F.~C. Adams, J.~Bond, K.~Freese, J.~A. Frieman and A.~V. Olinto, \emph{{Natural
  inflation: Particle physics models, power law spectra for large scale
  structure, and constraints from COBE}},
  \href{https://doi.org/10.1103/PhysRevD.47.426}{\emph{Phys. Rev. D} {\bfseries
  47} (1993) 426} [\href{https://arxiv.org/abs/hep-ph/9207245}{{\ttfamily
  hep-ph/9207245}}].

\bibitem{PhysRevD.98.030001}
{\scshape Particle Data Group} collaboration, M.~Tanabashi, K.~Hagiwara,
  K.~Hikasa, K.~Nakamura, Y.~Sumino, F.~Takahashi et~al., \emph{Review of
  particle physics},
  \href{https://doi.org/10.1103/PhysRevD.98.030001}{\emph{Phys. Rev. D}
  {\bfseries 98} (2018) 030001}.

\bibitem{Kaiser:2010ps}
D.~I. Kaiser, \emph{{Conformal Transformations with Multiple Scalar Fields}},
  \href{https://doi.org/10.1103/PhysRevD.81.084044}{\emph{Phys. Rev. D}
  {\bfseries 81} (2010) 084044}
  [\href{https://arxiv.org/abs/1003.1159}{{\ttfamily 1003.1159}}].

\bibitem{Abedi:2014mka}
H.~Abedi and A.~M. Abbassi, \emph{{Gravitational constant in multiple field
  gravity}}, \href{https://doi.org/10.1088/1475-7516/2015/05/026}{\emph{JCAP}
  {\bfseries 05} (2015) 026} [\href{https://arxiv.org/abs/1411.4854}{{\ttfamily
  1411.4854}}].

\bibitem{Achucarro:2016fby}
A.~Ach\'{u}carro, V.~Atal, C.~Germani and G.~A. Palma, \emph{{Cumulative
  effects in inflation with ultra-light entropy modes}},
  \href{https://doi.org/10.1088/1475-7516/2017/02/013}{\emph{JCAP} {\bfseries
  02} (2017) 013} [\href{https://arxiv.org/abs/1607.08609}{{\ttfamily
  1607.08609}}].

\bibitem{Bezrukov:2007ep}
F.~L. Bezrukov and M.~Shaposhnikov, \emph{{The Standard Model Higgs boson as
  the inflaton}},
  \href{https://doi.org/10.1016/j.physletb.2007.11.072}{\emph{Phys. Lett. B}
  {\bfseries 659} (2008) 703}
  [\href{https://arxiv.org/abs/0710.3755}{{\ttfamily 0710.3755}}].

\bibitem{Galante:2014ifa}
M.~Galante, R.~Kallosh, A.~Linde and D.~Roest, \emph{{Unity of Cosmological
  Inflation Attractors}},
  \href{https://doi.org/10.1103/PhysRevLett.114.141302}{\emph{Phys. Rev. Lett.}
  {\bfseries 114} (2015) 141302}
  [\href{https://arxiv.org/abs/1412.3797}{{\ttfamily 1412.3797}}].

\bibitem{Ferrara:2013rsa}
S.~Ferrara, R.~Kallosh, A.~Linde and M.~Porrati, \emph{{Minimal Supergravity
  Models of Inflation}},
  \href{https://doi.org/10.1103/PhysRevD.88.085038}{\emph{Phys. Rev. D}
  {\bfseries 88} (2013) 085038}
  [\href{https://arxiv.org/abs/1307.7696}{{\ttfamily 1307.7696}}].

\bibitem{Kallosh:2013yoa}
R.~Kallosh, A.~Linde and D.~Roest, \emph{{Superconformal Inflationary
  $\alpha$-Attractors}},
  \href{https://doi.org/10.1007/JHEP11(2013)198}{\emph{JHEP} {\bfseries 11}
  (2013) 198} [\href{https://arxiv.org/abs/1311.0472}{{\ttfamily 1311.0472}}].

\bibitem{Kallosh:2014rga}
R.~Kallosh, A.~Linde and D.~Roest, \emph{{Large field inflation and double
  $\alpha$-attractors}},
  \href{https://doi.org/10.1007/JHEP08(2014)052}{\emph{JHEP} {\bfseries 08}
  (2014) 052} [\href{https://arxiv.org/abs/1405.3646}{{\ttfamily 1405.3646}}].

\bibitem{Kallosh:2015lwa}
R.~Kallosh and A.~Linde, \emph{{Planck, LHC, and $\alpha$-attractors}},
  \href{https://doi.org/10.1103/PhysRevD.91.083528}{\emph{Phys. Rev. D}
  {\bfseries 91} (2015) 083528}
  [\href{https://arxiv.org/abs/1502.07733}{{\ttfamily 1502.07733}}].

\bibitem{Starobinsky:1980te}
A.~A. Starobinsky, \emph{{A New Type of Isotropic Cosmological Models Without
  Singularity}}, \href{https://doi.org/10.1016/0370-2693(80)90670-X}{\emph{Adv.
  Ser. Astrophys. Cosmol.} {\bfseries 3} (1987) 130}.

\bibitem{Kaiser:2013sna}
D.~I. Kaiser and E.~I. Sfakianakis, \emph{{Multifield Inflation after Planck:
  The Case for Nonminimal Couplings}},
  \href{https://doi.org/10.1103/PhysRevLett.112.011302}{\emph{Phys. Rev. Lett.}
  {\bfseries 112} (2014) 011302}
  [\href{https://arxiv.org/abs/1304.0363}{{\ttfamily 1304.0363}}].

\bibitem{Martin:2013tda}
J.~Martin, C.~Ringeval and V.~Vennin, \emph{{Encyclopaedia Inflationaris}},
  \href{https://doi.org/10.1016/j.dark.2014.01.003}{\emph{Phys. Dark Univ.}
  {\bfseries 5-6} (2014) 75} [\href{https://arxiv.org/abs/1303.3787}{{\ttfamily
  1303.3787}}].

\bibitem{Gordon:2000hv}
C.~Gordon, D.~Wands, B.~A. Bassett and R.~Maartens, \emph{{Adiabatic and
  entropy perturbations from inflation}},
  \href{https://doi.org/10.1103/PhysRevD.63.023506}{\emph{Phys. Rev.}
  {\bfseries D63} (2000) 023506}
  [\href{https://arxiv.org/abs/astro-ph/0009131}{{\ttfamily
  astro-ph/0009131}}].

\bibitem{Wands:2000dp}
D.~Wands, K.~A. Malik, D.~H. Lyth and A.~R. Liddle, \emph{{A New approach to
  the evolution of cosmological perturbations on large scales}},
  \href{https://doi.org/10.1103/PhysRevD.62.043527}{\emph{Phys. Rev. D}
  {\bfseries 62} (2000) 043527}
  [\href{https://arxiv.org/abs/astro-ph/0003278}{{\ttfamily
  astro-ph/0003278}}].

\bibitem{Senatore:2012ya}
L.~Senatore and M.~Zaldarriaga, \emph{{The constancy of $\zeta$ in single-clock
  Inflation at all loops}},
  \href{https://doi.org/10.1007/JHEP09(2013)148}{\emph{JHEP} {\bfseries 09}
  (2013) 148} [\href{https://arxiv.org/abs/1210.6048}{{\ttfamily 1210.6048}}].

\bibitem{Schutz:2013fua}
K.~Schutz, E.~I. Sfakianakis and D.~I. Kaiser, \emph{{Multifield Inflation
  after Planck: Isocurvature Modes from Nonminimal Couplings}},
  \href{https://doi.org/10.1103/PhysRevD.89.064044}{\emph{Phys. Rev. D}
  {\bfseries 89} (2014) 064044}
  [\href{https://arxiv.org/abs/1310.8285}{{\ttfamily 1310.8285}}].

\bibitem{DeCross:2015uza}
M.~P. DeCross, D.~I. Kaiser, A.~Prabhu, C.~Prescod-Weinstein and E.~I.
  Sfakianakis, \emph{{Preheating after Multifield Inflation with Nonminimal
  Couplings, I: Covariant Formalism and Attractor Behavior}},
  \href{https://doi.org/10.1103/PhysRevD.97.023526}{\emph{Phys. Rev. D}
  {\bfseries 97} (2018) 023526}
  [\href{https://arxiv.org/abs/1510.08553}{{\ttfamily 1510.08553}}].

\bibitem{DeCross:2016fdz}
M.~P. DeCross, D.~I. Kaiser, A.~Prabhu, C.~Prescod-Weinstein and E.~I.
  Sfakianakis, \emph{{Preheating after multifield inflation with nonminimal
  couplings, II: Resonance Structure}},
  \href{https://doi.org/10.1103/PhysRevD.97.023527}{\emph{Phys. Rev. D}
  {\bfseries 97} (2018) 023527}
  [\href{https://arxiv.org/abs/1610.08868}{{\ttfamily 1610.08868}}].

\bibitem{DeCross:2016cbs}
M.~P. DeCross, D.~I. Kaiser, A.~Prabhu, C.~Prescod-Weinstein and E.~I.
  Sfakianakis, \emph{{Preheating after multifield inflation with nonminimal
  couplings, III: Dynamical spacetime results}},
  \href{https://doi.org/10.1103/PhysRevD.97.023528}{\emph{Phys. Rev. D}
  {\bfseries 97} (2018) 023528}
  [\href{https://arxiv.org/abs/1610.08916}{{\ttfamily 1610.08916}}].

\bibitem{Nguyen:2019kbm}
R.~Nguyen, J.~van~de Vis, E.~I. Sfakianakis, J.~T. Giblin and D.~I. Kaiser,
  \emph{{Nonlinear Dynamics of Preheating after Multifield Inflation with
  Nonminimal Couplings}},
  \href{https://doi.org/10.1103/PhysRevLett.123.171301}{\emph{Phys. Rev. Lett.}
  {\bfseries 123} (2019) 171301}
  [\href{https://arxiv.org/abs/1905.12562}{{\ttfamily 1905.12562}}].

\bibitem{vandeVis:2020qcp}
J.~van~de Vis, R.~Nguyen, E.~I. Sfakianakis, J.~T. Giblin and D.~I. Kaiser,
  \emph{{Time-Scales for Nonlinear Processes in Preheating after Multifield
  Inflation with Nonminimal Couplings}},
  \href{https://arxiv.org/abs/2005.00433}{{\ttfamily 2005.00433}}.

\bibitem{Baumann:2009ds}
D.~Baumann, \emph{{Inflation}},  in \emph{{Theoretical Advanced Study Institute
  in Elementary Particle Physics}: {Physics of the Large and the Small}},
  pp.~523--686, 2011, \href{https://arxiv.org/abs/0907.5424}{{\ttfamily
  0907.5424}}, \href{https://doi.org/10.1142/9789814327183\_0010}{DOI}.

\bibitem{Ade:2018sbj}
{\scshape Simons Observatory} collaboration, P.~Ade et~al., \emph{{The Simons
  Observatory: Science goals and forecasts}},
  \href{https://doi.org/10.1088/1475-7516/2019/02/056}{\emph{JCAP} {\bfseries
  02} (2019) 056} [\href{https://arxiv.org/abs/1808.07445}{{\ttfamily
  1808.07445}}].

\bibitem{NANOGrav:2023hvm}
{\scshape NANOGrav} collaboration, A.~Afzal et~al., \emph{{The NANOGrav 15 yr
  Data Set: Search for Signals from New Physics}},
  \href{https://doi.org/10.3847/2041-8213/acdc91}{\emph{Astrophys. J. Lett.}
  {\bfseries 951} (2023) L11}
  [\href{https://arxiv.org/abs/2306.16219}{{\ttfamily 2306.16219}}].

\bibitem{Akrami:2018odb}
{\scshape Planck} collaboration, Y.~Akrami et~al., \emph{{Planck 2018 results.
  X. Constraints on inflation}},
  \href{https://arxiv.org/abs/1807.06211}{{\ttfamily 1807.06211}}.

\bibitem{Achucarro:2019pux}
A.~Ach\'{u}carro, E.~J. Copeland, O.~Iarygina, G.~A. Palma, D.-G. Wang and
  Y.~Welling, \emph{{Shift-Symmetric Orbital Inflation: single field or
  multi-field?}},  \href{https://arxiv.org/abs/1901.03657}{{\ttfamily
  1901.03657}}.

\bibitem{Achucarro:2019lgo}
A.~Ach\'{u}carro, G.~A. Palma, D.-G. Wang and Y.~Welling, \emph{{Origin of
  ultra-light fields during inflation and their suppressed non-Gaussianity}},
  \href{https://arxiv.org/abs/1908.06956}{{\ttfamily 1908.06956}}.

\bibitem{Achucarro:2019mea}
A.~Ach\'{u}carro and Y.~Welling, \emph{{Orbital Inflation: inflating along an
  angular isometry of field space}},
  \href{https://arxiv.org/abs/1907.02020}{{\ttfamily 1907.02020}}.

\bibitem{Chen:2014joa}
X.~Chen and M.~H. Namjoo, \emph{{Standard Clock in Primordial Density
  Perturbations and Cosmic Microwave Background}},
  \href{https://doi.org/10.1016/j.physletb.2014.11.002}{\emph{Phys. Lett. B}
  {\bfseries 739} (2014) 285}
  [\href{https://arxiv.org/abs/1404.1536}{{\ttfamily 1404.1536}}].

\bibitem{Chen:2014cwa}
X.~Chen, M.~H. Namjoo and Y.~Wang, \emph{{Models of the Primordial Standard
  Clock}}, \href{https://doi.org/10.1088/1475-7516/2015/02/027}{\emph{JCAP}
  {\bfseries 02} (2015) 027} [\href{https://arxiv.org/abs/1411.2349}{{\ttfamily
  1411.2349}}].

\bibitem{Chen:2015lza}
X.~Chen, M.~H. Namjoo and Y.~Wang, \emph{{Quantum Primordial Standard Clocks}},
  \href{https://doi.org/10.1088/1475-7516/2016/02/013}{\emph{JCAP} {\bfseries
  02} (2016) 013} [\href{https://arxiv.org/abs/1509.03930}{{\ttfamily
  1509.03930}}].

\bibitem{Chen:2016qce}
X.~Chen, M.~H. Namjoo and Y.~Wang, \emph{{A Direct Probe of the Evolutionary
  History of the Primordial Universe}},
  \href{https://doi.org/10.1007/s11433-016-0278-8}{\emph{Sci. China Phys. Mech.
  Astron.} {\bfseries 59} (2016) 101021}
  [\href{https://arxiv.org/abs/1608.01299}{{\ttfamily 1608.01299}}].

\bibitem{Domenech:2019cyh}
G.~Dom\`{e}nech and M.~Kamionkowski, \emph{{Lensing anomaly and oscillations in
  the primordial power spectrum}},
  \href{https://arxiv.org/abs/1905.04323}{{\ttfamily 1905.04323}}.

\bibitem{Braglia:2020fms}
M.~Braglia, D.~K. Hazra, L.~Sriramkumar and F.~Finelli, \emph{{Generating
  primordial features at large scales in two field models of inflation}},
  \href{https://arxiv.org/abs/2004.00672}{{\ttfamily 2004.00672}}.

\bibitem{Anber:2009ua}
M.~M. Anber and L.~Sorbo, \emph{{Naturally inflating on steep potentials
  through electromagnetic dissipation}},
  \href{https://doi.org/10.1103/PhysRevD.81.043534}{\emph{Phys.\ Rev.\ D}
  {\bfseries 81} (2010) 043534}
  [\href{https://arxiv.org/abs/0908.4089}{{\ttfamily 0908.4089}}].

\bibitem{Adshead:2015pva}
P.~Adshead, J.~T. Giblin, T.~R. Scully and E.~I. Sfakianakis,
  \emph{{Gauge-preheating and the end of axion inflation}},
  \href{https://doi.org/10.1088/1475-7516/2015/12/034}{\emph{JCAP} {\bfseries
  12} (2015) 034} [\href{https://arxiv.org/abs/1502.06506}{{\ttfamily
  1502.06506}}].

\bibitem{Adshead:2016iae}
P.~Adshead, J.~T. Giblin, T.~R. Scully and E.~I. Sfakianakis,
  \emph{{Magnetogenesis from axion inflation}},
  \href{https://doi.org/10.1088/1475-7516/2016/10/039}{\emph{JCAP} {\bfseries
  10} (2016) 039} [\href{https://arxiv.org/abs/1606.08474}{{\ttfamily
  1606.08474}}].

\bibitem{Garcia-Bellido:2016dkw}
J.~Garcia-Bellido, M.~Peloso and C.~Unal, \emph{{Gravitational waves at
  interferometer scales and primordial black holes in axion inflation}},
  \href{https://doi.org/10.1088/1475-7516/2016/12/031}{\emph{JCAP} {\bfseries
  12} (2016) 031} [\href{https://arxiv.org/abs/1610.03763}{{\ttfamily
  1610.03763}}].

\bibitem{Adshead:2019lbr}
P.~Adshead, J.~T. Giblin, M.~Pieroni and Z.~J. Weiner, \emph{{Constraining
  axion inflation with gravitational waves from preheating}},
  \href{https://doi.org/10.1103/PhysRevD.101.083534}{\emph{Phys. Rev. D}
  {\bfseries 101} (2020) 083534}
  [\href{https://arxiv.org/abs/1909.12842}{{\ttfamily 1909.12842}}].

\bibitem{Adshead:2019igv}
P.~Adshead, J.~T. Giblin, M.~Pieroni and Z.~J. Weiner, \emph{{Constraining
  axion inflation with gravitational waves across 29 decades in frequency}},
  \href{https://doi.org/10.1103/PhysRevLett.124.171301}{\emph{Phys. Rev. Lett.}
  {\bfseries 124} (2020) 171301}
  [\href{https://arxiv.org/abs/1909.12843}{{\ttfamily 1909.12843}}].

\bibitem{Lorenzoni:2024krn}
D.~L. Lorenzoni, D.~I. Kaiser and E.~McDonough, \emph{{Natural Inflation with
  Exponentially Small Tensor-To-Scalar Ratio}},
  \href{https://arxiv.org/abs/2405.13881}{{\ttfamily 2405.13881}}.

\bibitem{Adshead:2010mc}
P.~Adshead, R.~Easther, J.~Pritchard and A.~Loeb, \emph{{Inflation and the
  Scale Dependent Spectral Index: Prospects and Strategies}},
  \href{https://doi.org/10.1088/1475-7516/2011/02/021}{\emph{JCAP} {\bfseries
  02} (2011) 021} [\href{https://arxiv.org/abs/1007.3748}{{\ttfamily
  1007.3748}}].

\bibitem{Hardwick:2018zry}
R.~J. Hardwick, V.~Vennin and D.~Wands, \emph{{The decisive future of
  inflation}}, \href{https://doi.org/10.1088/1475-7516/2018/05/070}{\emph{JCAP}
  {\bfseries 05} (2018) 070}
  [\href{https://arxiv.org/abs/1803.09491}{{\ttfamily 1803.09491}}].

\bibitem{Easther:2021rdg}
R.~Easther, B.~Bahr-Kalus and D.~Parkinson, \emph{{Running primordial
  perturbations: Inflationary dynamics and observational constraints}},
  \href{https://doi.org/10.1103/PhysRevD.106.L061301}{\emph{Phys. Rev. D}
  {\bfseries 106} (2022) L061301}
  [\href{https://arxiv.org/abs/2112.10922}{{\ttfamily 2112.10922}}].

\bibitem{Bahr-Kalus:2022prj}
B.~Bahr-Kalus, D.~Parkinson and R.~Easther, \emph{{Constraining cosmic
  inflation with observations: Prospects for 2030}},
  \href{https://doi.org/10.1093/mnras/stad092}{\emph{Mon. Not. Roy. Astron.
  Soc.} {\bfseries 520} (2023) 2405}
  [\href{https://arxiv.org/abs/2212.04115}{{\ttfamily 2212.04115}}].

\bibitem{Martin:2024nlo}
J.~Martin, C.~Ringeval and V.~Vennin, \emph{{Vanilla Inflation Predicts
  Negative Running}},  \href{https://arxiv.org/abs/2404.15089}{{\ttfamily
  2404.15089}}.

\bibitem{Munoz:2016owz}
J.~B. Mu\~noz, E.~D. Kovetz, A.~Raccanelli, M.~Kamionkowski and J.~Silk,
  \emph{{Towards a measurement of the spectral runnings}},
  \href{https://doi.org/10.1088/1475-7516/2017/05/032}{\emph{JCAP} {\bfseries
  05} (2017) 032} [\href{https://arxiv.org/abs/1611.05883}{{\ttfamily
  1611.05883}}].

\bibitem{Chluba:2019kpb}
J.~Chluba et~al., \emph{{Spectral Distortions of the CMB as a Probe of
  Inflation, Recombination, Structure Formation and Particle Physics}:
  {Astro2020 Science White Paper}}, {\emph{Bull. Am. Astron. Soc.} {\bfseries
  51} (2019) 184} [\href{https://arxiv.org/abs/1903.04218}{{\ttfamily
  1903.04218}}].

\bibitem{Josan:2009qn}
A.~S. Josan, A.~M. Green and K.~A. Malik, \emph{{Generalised constraints on the
  curvature perturbation from primordial black holes}},
  \href{https://doi.org/10.1103/PhysRevD.79.103520}{\emph{Phys. Rev. D}
  {\bfseries 79} (2009) 103520}
  [\href{https://arxiv.org/abs/0903.3184}{{\ttfamily 0903.3184}}].

\bibitem{Iarygina:2023mtj}
O.~Iarygina, E.~I. Sfakianakis, R.~Sharma and A.~Brandenburg,
  \emph{{Backreaction of axion-SU(2) dynamics during inflation}},
  \href{https://arxiv.org/abs/2311.07557}{{\ttfamily 2311.07557}}.

\bibitem{McDonough:2016xvu}
E.~McDonough, H.~Bazrafshan~Moghaddam and R.~H. Brandenberger,
  \emph{{Preheating and Entropy Perturbations in Axion Monodromy Inflation}},
  \href{https://doi.org/10.1088/1475-7516/2016/05/012}{\emph{JCAP} {\bfseries
  05} (2016) 012} [\href{https://arxiv.org/abs/1601.07749}{{\ttfamily
  1601.07749}}].

\bibitem{Lozanov:2019jff}
K.~D. Lozanov and M.~A. Amin, \emph{{GFiRe: a Gauge Field integrator for
  Reheating}}, \href{https://doi.org/10.1088/1475-7516/2020/04/058}{\emph{JCAP}
  {\bfseries 04} (2020) 058}
  [\href{https://arxiv.org/abs/1911.06827}{{\ttfamily 1911.06827}}].

\end{thebibliography}\endgroup

\end{document}